\documentclass[12pt]{article}
\pdfoutput=1
\usepackage{mathrsfs}
\usepackage{natbib,graphicx,setspace,lscape,longtable,xr}
\usepackage{mathrsfs,amsmath,amsthm,amssymb,color}
\usepackage{natbib,epsfig,graphicx,pdfpages}
\usepackage{rotating}
\usepackage{float}
\usepackage{bm}
\usepackage{ulem}
\usepackage{CJK}
\usepackage{ragged2e}
\usepackage{setspace}
\usepackage{threeparttable}
\usepackage{multirow}
\usepackage{enumitem}
\usepackage{hyperref}
\usepackage{etoolbox}
\usepackage{subfig}
\usepackage[ruled]{algorithm2e}
\usepackage{adjustbox}
\usepackage{blindtext}
\usepackage{authblk}
\usepackage{enumerate}
\usepackage{cleveref}
\usepackage{mathtools}
\usepackage[titletoc,title]{appendix}
\usepackage{booktabs}
\usepackage{diagbox}

\bibpunct{(}{)}{;}{a}{,}{,}

\newtheorem{theorem}{Theorem}
\newtheorem{lemma}{Lemma}
\newtheorem{proposition}{Proposition}

\allowdisplaybreaks[4]

\newcounter{Condition}
\newcommand{\mylabel}[2]{#2\def\@currentlabel{#2}\label{#1}}
\newcommand{\csection}[1]
{\begin{center}
		\stepcounter{section}
		{\bf\large\arabic{section}. #1}
	\end{center}
}

\newcommand{\scsection}[1]
{\begin{center}
		{\bf\large #1}
	\end{center}
}

\newcommand{\csubsection}[1]{
	\begin{center}
		\stepcounter{subsection}
		{\it\arabic{section}.\arabic{subsection}. #1}
	\end{center}
}

\newcommand{\scsubsection}[1]{
	\begin{center}
		\stepcounter{subsection}
		{\it #1}
	\end{center}
}

\def\tr{\mbox{tr}}

\def\ve{\varepsilon}

\def\beq{\begin{equation}}
	\def\eeq{\end{equation}}
\def\beqr{\begin{eqnarray}}
	\def\eeqr{\end{eqnarray}}
\def\beqrs{\begin{eqnarray*}}
	\def\eeqrs{\end{eqnarray*}}
\def\bet{\begin{theorem}}
	\def\eet{\end{theorem}}
\def\bel{\begin{lemma}}
	\def\eel{\end{lemma}}
\def\bep{\begin{proposition}}
	\def\eep{\end{proposition}}
\def\bg{\begin{figure}[tbph]\begin{center}}
		\def\eg{\end{center}\end{figure}}

\def\bc{\begin{center}}
	\def\ec{\end{center}}

\def\wt{\widetilde}

\def\wh{\widehat}

\def\mB{\mathbb B}

\def\mI{\mathbb I}

\def\mR{\mathbb{R}}
\def\mS{\mathbb S}
\def\mL{\mathcal L}

\def\wmH{\widehat{\mathbb H}}

\def\mF{\mathcal F}

\def\mS{\mathcal S}

\def\mW{\mathbb{W}}

\def\mX{\mathbb{X}}

\def\mZ{\mathbb{Z}}
\def\var{\mbox{var}}

\def\argmax{\mbox{argmax}}

\def\diag{\mbox{diag}}

\renewcommand{\arraystretch}{1.3}

\usepackage[a4paper,bindingoffset=0.2in,%
left=1in,right=1in,top=1in,bottom=1in,%
footskip=.25in]{geometry}

\def\boxit#1{\vbox{\hrule\hbox{\vrule\kern6pt\vbox{\kern6pt#1\kern6pt}\kern6pt\vrule}\hrule}}

\numberwithin{equation}{section}

\begin{document}

\begin{center}
{\bf\Large Testing Sufficiency for Transfer Learning}

Ziqian Lin$^{1}$, Yuan Gao$^{2}$, Feifei Wang$^{3,4}$\footnote{The corresponding author. Email: feifei.wang@ruc.edu.cn}, and Hansheng Wang$^1$

{\it\small
$^1$Guanghua School of Management, Peking University, Beijing, China;
$^2$School of Statistics and KLATASDS-MOE, East China Normal University, Shanghai, China;
$^3$Center for Applied Statistics, Renmin University of China, Beijing, China;
$^4$School of Statistics, Renmin University of China, Beijing, China.
}

\end{center}

\begin{singlespace}
\begin{abstract}
Modern statistical analysis often encounters high dimensional models but with limited sample sizes. This makes the target data based statistical estimation very difficult. Then how to borrow information from another large sized source data for more accurate target model estimation becomes an interesting problem. This leads to the useful idea of transfer learning. Various estimation methods in this regard have been developed recently. In this work, we study transfer learning from a different perspective. Specifically, we consider here the problem of testing for transfer learning sufficiency. By transfer learning sufficiency (denoted as the null hypothesis), we mean that, with the help of the source data, the useful information contained in the feature vectors of the target data can be sufficiently extracted for predicting the interested target response. Therefore, the rejection of the null hypothesis implies that information useful for prediction remains in the feature vectors of the target data and thus calls for further exploration. To this end, we develop a novel testing procedure and a centralized and standardized test statistic, whose asymptotic null distribution is analytically derived. Simulation studies are presented to demonstrate the finite sample performance of the proposed method. A deep learning related real data example is presented for illustration purpose.
\end{abstract}


\end{singlespace}
\clearpage

\csection{INTRODUCTION}

Modern statistical analysis often encounters challenging situations, where the target model is of high dimension but the accessible sample is of a very limited size. This is particularly true for medical, genetic, and image related studies, where the supporting samples are hard to obtain due to ethical or cost issues \citep{li2020transfer}. Consider for example the analysis of skin cutaneous melanoma (SKCM) disease. The widely used SKCM dataset\footnote{It is available from the Cancer Genome Atlas (TCGA) by https://www.cancer.gov/about-nci/organization/ccg/research/structural-genomics/tcga} consists of only 347 patient samples but a total of 18,335 gene expression levels measured for each sample. Consider modern deep learning researches as another example. For this research field, the topic of few-shot learning (i.e., training models on a few samples) has attracted arising attentions from researchers, because data collection and labeling is often very expensive \citep{10.1145/3386252}. Given the fact that deep learning models often have thousands of hundreds of parameters to estimate, the problem of ``small sample size and high dimensionality" becomes inevitable. Then, how to accurately train a high dimensional target model with a limited sample size becomes a practically important problem.

One seemingly very promising way to solve this problem is transfer learning \citep{pan2009survey}. By transfer learning, we expect that there exists another huge sized auxiliary dataset, which is different from but somewhat related with the target dataset. For convenience, we refer to this auxiliary dataset as a source dataset. The probability distribution generating the source/target dataset is referred to as the source/target population. It is remarkable that the source population is often different from the target population. To fix this idea, let ($X,Y$) and ($X^*,Y^*$) be the observations generated from the target population and source population, respectively. The interested response $Y$ in the target population could be totally different from $Y^*$ in the source population. For example, $Y^*$ is often assumed to be a categorical response taken values in $\{0,1,...,K\}$, but $Y$ is allowed to be continuous. Nevertheless, the covariate $X$ and $X^*$ must be of the same dimension $p$. Otherwise, no transfer learning can be conducted. Although the sample size of target data is often limited, the size of the source data is expected to be huge. Consequently, various high dimensional models can be estimated on the source data with reasonable accuracy. Then how to transfer the valuable information learned from the source data to the target model becomes the key issue. To this end, various transfer learning methods have been developed.

In the computer science literature, there exist a variety of strategies for transfer learning; see \cite{Weiss2016A} and \cite{Farahani2020A} for an excellent review. Among them, the arguably most typical transfer learning strategy is the feature-based approach \citep{6909618,shin2016deep,guo2018deep}. The key idea is to find a common feature space underlying the source data and having predictive power for the target model in the meanwhile. Consider for example a categorical response $Y^*$. Then a deep learning model can often be written as $P(Y^*=k)=\exp\{Z_{\mathbb B}^{\top}(X^{*})\gamma_k\}/1+\sum_{k=1}^K\exp\{Z_{\mathbb B}^{\top}(X^{*})\gamma_k\}$, where $Z_{\mathbb B}(\cdot)$ is a feature generating function \citep{donahue2014decaf,he2016deep,2021Multi}. It transforms the high dimensional vector $X^*$ into a lower dimensional one as $Z_{\mathbb B}(X^{*})$ in the feature space. Since the dimension of $X^*$ is large, the dimension of the unknown parameter $\mathbb B$ for the feature generating function is also high. Fortunately, the source data size is often huge. Therefore, the unknown parameter $\mathbb B$ (or its linear transformation) can be estimated with reasonable accuracy. Once $\mathbb B$ is estimated (denoted by $\widehat{\mathbb B}$), the estimated feature generating function $Z_{\widehat{\mathbb B}}(\cdot)$ can be readily applied to the target data with the high dimensional covariate $X$. This leads to the newly estimated feature vector $Z_{\widehat{\mathbb B}}(X)$. By modeling the relationship between $Z_{\widehat{\mathbb B}}(X)$ and $Y$, the valuable information learned from the source data can be transferred to the target model.

Recently, transfer learning has attracted great attention from the statistical literature, but from a somewhat different perspective. A number of researchers have developed novel new algorithms for different models, such as linear regression \citep{li2020transfer}, generalized linear regression \citep{Tian2022Transfer}, nonparametric regression \citep{Cai2021Transfer}, multiple testing \citep{Liang2022locally}, and Gaussian graphical models \citep{Li2022Transfer}. The general idea here is to estimate the target model and source model simultaneously, and then utilize the similarity between the source and target models to improve the estimation accuracy of the target model. For example, \cite{li2020transfer} propose a Trans-Lasso method for transfer learning in a high dimensional linear regression model. The basic idea is to pool source data and target data together so that an initial estimator can be obtained. Thereafter, the bias of the initial estimator is corrected by the target data. Similar methods are also developed for generalized linear regression models by \cite{Tian2022Transfer}.

Inspired by these pioneer researches from both the computer science literature and statistical literature, we take here a very unique perspective to study transfer learning. On one side, our approach shares similar spirit as those in the computer science literature. Specifically, we assume a feature generating function $Z_{\mathbb B}(X)$, which can be learned by the source data and then transferred to the target model for generating lower dimensional feature vectors. We follow this direction mainly because it is probably the most widely adopted transfer learning strategy in practice. In fact, many most commonly used deep learning frameworks (e.g., TensorFlow and Pytorch) have implemented this strategy for transfer learning. On the other side, our proposed method also benefits from those pioneered works in the statistical literature. We are inspired by the statistical theory developed therein. By these inspirational statistical theory, one can gain better understanding about the information transferring dynamics between the source data and target data. As a consequence, new estimation and inference procedure can be developed with statistical guarantees.

As our first attempt, we follow the past literature to start with the simplest but possibly the most important feature generating function, i.e., the linear transformation \citep{Lu2010On,Fan2013Linear}. Specifically, we assume for the source data a standard multinominal logistic regression model as
$P(Y^*=k)=\exp\{X^{*\top}\beta_k\}/[1+\sum_{k=1}^K\exp\{X^{*\top}\beta_k\}]$ for $1\leq k \leq K$. Then the arguably simplest feature generating function can be defined as $Z^*=Z_{\mathbb B}(X^*)=\mathbb{B}^{\top}X^{*}$, where $\mathbb{B}=(\beta_1,\beta_2,...,\beta_K)\in \mR^{p\times K}$ with $K\ll p$. We next assume for the target data a binary logistic regression model as
$P(Y=1)=\exp(X^{\top}\theta)/\{1+\exp(X^{\top}\theta)\}$, where $\theta \in \mR^p$ is the target parameter of interest. We then consider how to transfer the knowledge learned from the source data about $\mathbb B$ to help the estimation of $\theta$. The key issue here is whether $\theta \in \mathcal{S}(\mathbb B)$, where $\mathcal{S}(\mathbb B)$ stands for the linear space generated by the column vectors of $\mathbb B$. If $\theta \in \mathcal{S}(\mathbb B)$ indeed happens, we then should have $\theta=\mathbb B \gamma$ for some $\gamma \in \mR^{K}$. Accordingly, the target model can be rewritten as $P(Y=1)=\exp(Z^{\top}\gamma)/\{1+\exp(Z^{\top}\gamma)\}$, where $Z =\mathbb{B}^{\top}X \in \mR^{K}$. Thereafter, instead of working on the high dimensional vector $X$ directly, we can work on the dimension-reduced vector $Z$ without any loss about the prediction accuracy. In this case, we say the transfer learning is sufficient in the sense that $P(Y=1|\mathbb{B}^{\top}X)=P(Y=1|X)$. Otherwise, if $\theta \notin \mathcal{S}(\mathbb B)$, we should have $P(Y=1|\mathbb{B}^{\top}X)\neq P(Y=1|X)$ for a significant amount of $X$. Therefore, there exists non-negligible loss in prediction accuracy. In this case, we say the transfer learning is insufficient. In other words, valuable information useful for more accurate prediction about $Y$ remains in $X$ and has not been fully captured by $\mathbb{B}^{\top}X$.

Practically, $\mathbb{B}$ is an unknown parameter. Then how to estimate $\mathbb{B}$ becomes an important problem. With the help of source data, we can easily obtain an estimate $\widehat{\mathbb{B}}=(\widehat{\beta}_1,\widehat{\beta}_2,...,\widehat{\beta}_K)$, where $\widehat{\beta}_k$ is (for example) the maximum likelihood estimator of $\beta_k$ computed on the source data. Under appropriate regularity conditions, we should have $\widehat{\mathbb{B}}$ to be consistent and asymptotically normal as $N\to\infty$. Since the source data usually have a huge sample size, we can expect $\widehat{\mathbb{B}}$ to have sufficient accuracy. With the estimated $\widehat{\mathbb{B}}$, the lower dimensional vector $Z$ can be replaced by $\widehat{Z}=\widehat{\mathbb{B}}^{\top}X$. Then the target model using $\widehat{Z}$ as the input can be estimated accordingly. This leads to a maximum likelihood estimator $\widehat{\gamma}$ for the coefficient $\gamma$. With the relationship $\theta=\mathbb B\gamma$, an estimator for $\theta$ can be defined as $\widehat{\theta} = \widehat{\mathbb{B}}\widehat \gamma$, which we refer to as the transfer learning (TL) estimator. To theoretically support our method, a rigorous asymptotic theory has been developed. We find that, under appropriate regularity conditions, the resulting TL estimator $\widehat{\theta}$ can be asymptotically as efficient as the oracle estimator. Here the oracle estimator is defined to be the ideal estimator obtained with a perfectly recovered lower dimensional feature vector $Z_{\mathbb B}(X)$.

Since the target data are often extremely valuable, we wish the useful information contained in the target feature has been fully utilized with the help of transfer learning. Otherwise, a significant amount of valuable information remains in the target feature for further exploration. This leads to an interesting theoretical problem. That is how to test sufficiency for transfer learning. To solve the problem, we develop here a novel testing procedure. The key idea is as follows. As mentioned before, a consistent estimator for the regression coefficient $\theta$ can be obtained for the target model, under the null hypothesis of transfer learning sufficiency and with the help of the source data. Note that this task can hardly be accomplished by using the target data only. Thereafter, pseudo residuals can be differentiated. Under appropriate regularity conditions, they should be uncorrelated with every single feature asymptotically. Accordingly, we should expect the sample covariance between the pseudo residual and each predictor to be close to 0. We are then inspired to have those sample covariances computed for each predictor squared and summed up together. This leads to an interesting test statistic sharing similar spirit as that of \cite{lan2014testing}. However, as demonstrated by \cite{lan2014testing}, to make test statistic of this form have a non-degenerate asymptotic distribution, appropriate centralization operation is necessarily needed. This leads to a centralized test statistic with significantly reduced bias and variability. We show rigorously that this centralized test statistic is asymptotically distributed as a standard normal distribution under the null hypothesis and after appropriate location-scale transformation. Extensive simulation studies are presented to demonstrate this testing procedure's finite sample performance. A deep learning related real dataset is analyzed for illustration purpose.

The rest of the article is organized as follows. Section 2 describes our theoretical framework for transfer learning. Based on the proposed theoretical framework, a transfer-learning based maximum likelihood estimator is developed for the target model. Its asymptotical properties are then rigorously studied. Thereafter, the problem of testing sufficiency for transfer learning is studied. In this regard, a high dimensional testing procedure is developed based on the sum of squared sample covariances between the pseudo-residuals and predictors. After appropriate centralization and standardization, its asymptotical null distribution is rigorously established. The finite sample performance of the proposed methodology is then evaluated by extensive numerical studies on both simulated and real datasets in Section 3. Lastly, Section 4 concludes the paper with a brief discussion.

\csection{TRANSFER LEARNING MODEL}
\csubsection{Transfer Learning Estimator}

Consider a target dataset with a total of $n$ samples. Let $(X_i,Y_i)$ be the observation collected from the $i$th subject ($1\le i\le n $) in the target dataset. Here $Y_i\in \{0,1\}$ is a binary response and $X_i = (X_{i1}, \dots, X_{ip})^\top \in \mR^p$ is the associated $p$-dimensional features. Without loss of generality, we assume all features in $X_i$ are centralized such that $E(X_i) = 0$. Further assume that $(X_i,Y_i)$ for different subjects are independently and identically distributed. To model the regression relationship between $X_i$ and $Y_i$, we consider a standard logistic regression model as
\beq
P\Big(Y_i=1\big|X_i\Big) = \frac{\exp(X_i^\top\theta)}{1+\exp(X_i^\top\theta)} = g\Big(X_i^\top \theta\Big), \label{logistic model}
\eeq
where $\theta = (\theta_1,\dots,\theta_p)^{\top}\in\mR^p$ is the $p$-dimensional regression coefficient, and $g(x) = \exp(x) / \{1+\exp(x)\}$ is the sigmoid function for any scalar $x$. We refer to \eqref{logistic model} as the target model. To estimate the target model, the maximum likelihood estimation method can be used. Specifically, a log-likelihood function can be constructed as
\beq
\mL_n^{\text{Target}}(\theta) = \sum_{i=1}^n \Bigg[Y_i \Big(X_i^\top\theta\Big)  - \log\bigg\{1+ \exp\Big(X_i^\top\theta\Big) \bigg\}\Bigg]. \label{target likelihood}
\eeq
Then the maximum likelihood estimator (MLE) of the target model can be obtained as $\widehat{\theta}_{\text{mle}}^{\text{Target} }= \argmax_{\theta}\mL_n^{\text{Target}}(\theta)$. If the feature dimension $p$ is fixed and the target sample size $n\to\infty$, we should have $\widehat{\theta}_{\text{mle}}^{\text{Target}}$ to be $\sqrt{n}$-consistent and asymptotically normal \citep{McCullagh1989,shao2003mathematical}. Unfortunately, we often encounter the situation, where the target sample size $n$ is very limited but the feature dimension $p$ is ultrahigh. This leads to a challenging situation with $n\approx p$ or $n<p$. In this case, the maximum likelihood estimator should be biased or even computationally infeasible \citep{Sur2019A,EJ2020THE}.

To solve this problem, we can seek for the help of transfer learning.
The general idea is to borrow information from the source data. Specifically, define $(X_i^*,Y_i^*)$ to be the observation collected for the $i$th sample with $1\le i \le N$ in the source data. Assume the covariate $X_i^* = (X_{i1}^*, \dots, X_{ip}^*)^\top \in\mR^p$ to have the same dimension as $X_i$. Similarly, we also assume $E(X_i^*) = 0$. However, different from the target data, we assume the response $Y_i^*$ in the source data to be a $(K+1)$-level categorical variable, i.e., $Y_i^* \in \{0,1,\dots,K \}$. We typically expect $K$ to be relatively large so that ample amount of information can be provided by $Y_i^*$. In the meanwhile, $K$ should not be too large neither. Instead, it should be much smaller than the target sample size $n$. Otherwise, the rich information provided by $Y_i^*$ cannot be conveniently transferred to the target model with a limited sample size.

To study the regression relationship between $X_i^*$ and $Y_i^*$, we assume the following multi-class logistic regression model as
\beq
P(Y_i^* = k | X_i^*) = \frac{\exp(X_i^{*\top}\beta_k)}{1+\sum_{j=1}^{K}\exp(X_i^{*\top}\beta_j)} \text{ for } 1\le k\le K, \label{source model}
\eeq
where $\beta_k = (\beta_{k1}, \dots, \beta_{kp})^\top\in\mR^p$ is the $p$-dimension regression coefficient for class $k$ with respect to the base class $0$. We refer to \eqref{source model} as the source model, and define $\mathbb{B} = (\beta_1,\dots,\beta_K) \in \mR^{p \times K}$ to be the corresponding coefficient matrix. Then the log-likelihood function of the source model can be spelled out as
\[
\mL^{\text{Source}}_N(\mathbb{B}) = \sum_{i=1}^{N}\Bigg[\sum_{k=1}^{K} I(Y_i^*  = k)X_i^{*\top}\beta_k - \log\bigg\{1+\sum_{k=1}^K \exp\Big(X_i^{*\top}\beta_k\Big)\bigg\} \Bigg].
\]
Thereafter, the maximum likelihood estimator of the source model can be computed as $\widehat{\mathbb{B}}^{\text{Source}}_{\text{mle}} = (\widehat\beta_1,\dots,\widehat\beta_K) =  \argmax_{\mathbb{B}} \mL^{\text{Source}}_N(\mathbb{B})$. Under appropriate regularity conditions, we should have $\widehat{\mathbb{B}}^{\text{Source}}_{\text{mle}}$ to be consistent and asymptotically normal as $N\to\infty$ \citep{fan2004nonconcave}. It is remarkable that, we assume in this work the source sample size $N$ is much larger than both the target sample size $n$ and the feature dimension $p$. Therefore, based on the large source data, the parameter $\mathbb{B}$ can be estimated with sufficient accuracy. Next, we consider how to estimate the parameter $\theta$ for the target model by transferring useful information from $\widehat{\mathbb{B}}^{\text{Source}}_{\text{mle}}$.

To borrow information from the source data by $\widehat{\mathbb{B}}^{\text{Source}}_{\text{mle}}$, we need to define a feature generating function $Z_{\mathbb B}(X)$, which can be learned by the source data and then transferred to the target data. Given the source model \eqref{source model}, we can define an arguably simplest
feature generating function as $Z^*_i = Z_{\mathbb B}(X^*_i) = \mathbb{B}^\top X^*_i \in \mR^K$ for the source data. We typically assume $K\ll p$, so that the dimension of the new feature $Z^*_i$ generated by the feature generating function is much smaller than that of the original feature $X^*_i$. Then the question is whether this feature generating function $Z_{\mathbb B}(X)$ is applicable for the target data. In other words, whether the rich information contained in the original high dimensional feature $X_i$ in the target data can be sufficiently represented by the dimension-reduced feature $Z_i = Z_{\mathbb B}(X_i) = \mathbb{B}^\top X_i \in \mR^K$. If the answer is positive, we should be able to write $X_i^{\top}\theta=
Z_{\mathbb B}^{\top}(X_i)\gamma = (X_i^{\top}\mathbb{B})\gamma=Z_i^{\top}\gamma$ for some $\gamma = (\gamma_1,\dots,\gamma_K)^\top \in \mR^K $. Accordingly, we should have $\theta=\mathbb{B}\gamma$. If this happens, we should have $P(Y_i=1|X_i)=P(Y_i=1|Z_i)$. Therefore, we say the transfer learning offered by $Z_{\mathbb B}(X)$ is sufficient. Otherwise, we say the transfer learning is insufficient.

Recall that the dimension of $Z_i$ is much smaller than the dimension of $X_i$. Therefore, instead of working on the high dimensional vector $X_i$ directly, we can work on the dimension-reduced vector $Z_i$ with the regression coefficient $\gamma$. Then the target model \eqref{logistic model} can be rewritten as
\[
P\Big(Y_i = 1\big|Z_i\Big) = \frac{\exp(Z_i^\top\gamma)}{1+\exp(Z_i^\top\gamma)}. \label{oracle model}
\]
Since $\mathbb{B}$ is an unknown parameter, we have to replace it by some appropriate estimator. As discussed before, one natural choice is the maximum likelihood estimator $\widehat{\mathbb{B}}_{\text{mle}}^{\text{Source}}$. Accordingly, we can replace $Z_i$ by $\widehat Z_i = \widehat{\mathbb{B}}_{\text{mle}}^{\text{Source}\top}X_i$. This leads to the following working model as
\[
P\Big(Y_i = 1\big| Z_i\Big) \approx \frac{\exp(\widehat{Z}_i^\top\gamma)}{1+\exp(\widehat{Z}_i^\top\gamma)}.
\]
Accordingly, a working log-likelihood function can be written as
\beq
\mL_n(\gamma;\widehat Z) = \sum_{i=1}^n \Bigg[Y_i \Big(\widehat Z_i^\top\gamma\Big)  - \log\bigg\{1+ \exp\Big(\widehat Z_i^\top\gamma\Big) \bigg\}\Bigg]. \label{working likelihood}
\eeq
Then a maximum likelihood type estimator for $\gamma$ can be obtained by maximizing \eqref{working likelihood} as $\widehat\gamma = \argmax_{\gamma} {\mL}_n(\gamma;\widehat Z)$. Recall that $\theta = \mathbb{B} \gamma$. Then an estimator for $\theta$ can be defined as $\widehat{\theta}^{\text{Transfer}} = \widehat{\mathbb{B}}_{\text{mle}}^{\text{Source}} \widehat \gamma$, which utilizes the information not only learned from the target data but also transferred from the source data. For convenience, we refer to it as a transfer learning (TL) estimator.

\newpage
\csubsection{The Transferred Maximum Likelihood Estimator}

We next study the asymptotic properties of the TL estimator. Let $\lambda_{\min}(A)$ and $\lambda_{\max}(A)$ be the smallest and largest eigenvalues of an arbitrary square matrix $A$, respectively. Define $\pi_{ik} = P(Y_i^* = k|X_i^*)$ and $\pi_i = (\pi_{i1}, \dots, \pi_{iK})^\top \in \mR^K$. Write $W_i = \diag(\pi_i) - \pi_i\pi_i^\top$. Define $\dot\mL_n(\gamma;Z)\in \mR^K$ to be the first-order derivative of the oracle log-likelihood function $\mL_n(\gamma;Z)$ with respect to $\gamma$. Here the oracle log-likelihood function $\mL_n(\gamma;Z)$ is computed based on the true feature vector $Z_i$ instead of its estimated counterpart $\widehat{Z}_i$. Further define $I(\gamma) = n^{-1}E\big[\dot\mL_n(\gamma;Z)\dot\mL_n(\gamma;Z)^\top\big] = E\big[g(Z_i^\top\gamma)\{1-g(Z_i^\top\gamma)\}Z_iZ_i^\top\big]$ as the Fisher information matrix of the oracle log-likelihood function. Next, we define the sub-Gaussian norm for a random variable $U$ as $ \|U\|_{\psi_2} = \inf\{t>0: E\exp(U^2/t^2)\le 2\}$, and for a random vector $V$ as $\|V\|_{\psi_2} = \sup_{\|u\| = 1}\|u^\top V\|_{\psi_2}$. Then we say a random vector $V$ is sub-Gaussian if $\|V\|_{\psi_2} <\infty $; see \cite{vershynin2018high} for more detailed discussions about the sub-Gaussian norm. For an arbitrary matrix $A = (a_{ij}) \in \mR^{m\times n}$, define its Frobenius norm as $\|A\|_F = \sqrt{\sum_{i=1}^m\sum_{j=1}^n a_{ij}^2}$. To study the theoretical properties of the TL estimator, the following technical conditions are necessarily needed:
\begin{itemize}
\item[(C1)] As $N\to\infty$, we should have $n^2 \log N/ N \to 0$ and $p \to \infty$. Moreover, there exists a positive constant $0<C<\infty $ such that $p\le Cn$.
\item[(C2)] Both the random vectors $X_i^*$ and $X_i$ are sub-Gaussian.\label{C2}
\item[(C3)] Denote $\otimes$ to be the kronecker product. As $p \to \infty$, there exist two positive constants $0<C_{\min} \le C_{\max}<\infty$ such that: (1) $C_{\min}\le \lambda_{\min}\{E (X_i^*X_i^{*\top}) \}\le \lambda_{\max}\{E (X_i^*X_i^{*\top}) \} \le C_{\max}$, (2) $C_{\min}\le \lambda_{\min}[E\{W_i\otimes (X_i^*X_i^{*\top })\}] \le \lambda_{\max}[E\{W_i\otimes (X_i^*X_i^{*\top })\}] \le C_{\max} $, (3) $C_{\min} \le \lambda_{\min}\{E(X_iX_i^{\top })\} \le\lambda_{\max}\{E (X_iX_i^{\top}) \} \le C_{\max}$, and (4) $C_{\min} \le  \lambda_{\min}(\Sigma_\gamma) \le \lambda_{\max}(\Sigma_\gamma) \le C_{\max}$. Here $\Sigma_\gamma = E(\ve_i^2 X_i X_i^\top)$, where $\ve_i=Y_i-E(Y_i|X_i)$ is the pseudo residual. \label{C3}
\item[(C4)] As $p\to\infty$, there exists a positive constant $\tau_{\max}>0$ such that $E\|Z_i\|^4 \le \tau_{\max}$ and $E(X_i^\top X_j)^4 \le \tau_{\max}p^2$ for any $i\neq j$.\label{C4}
\item[(C5)] As $p\to\infty$, we have $\lambda_{\max}(\mathbb B^\top \mathbb B) \le \tau_{\max}$, where $\tau_{\max}$ is the same as in (C4).
\end{itemize}
\noindent Condition (C1) implies that, the sample size of the source data should be sufficiently large in the sense that $n^2\log N/N\to 0$. Condition (C1) also restricts the divergence rate of the feature dimension $p$ such that it should not be of higher order than the target sample size $n$. However, we allow $p>n$. In that case, we should have $C>1$. Condition (C2) is slightly stronger than a standard moment condition. It is trivially satisfied if $X_i$ follows a multivariate normal distribution. A similar but stronger condition as $\sup_{i}\|X_i\| = O_p(\sqrt p)$ has been used in the past literature \citep{portnoy1985asymptotic,welsh1989m,he2000parameters,wang2011gee}. Condition (C3) imposes uniform bounds on the maximum and minimum eigenvalues of various positive definite matrices with diverging dimension $p$. Condition (C4) imposes two moment conditions about $Z_i$ and $X_i$, which are fairly reasonable. To fix the idea, consider for example a special case with $X_i$ following a multivariate standard normal distribution. Then it can be verified that $E(X_i^\top X_j)^4 = 3\tr^2\{E(X_iX_i^\top)\} + 6\tr\{E^4(X_iX_i^\top)\} = 3p^2 + 6p \le \tau_{\max} p^2$ with $\tau_{\max} = 9$. Condition (C5) is also a fairly reasonable assumption. To gain some quick understanding about this condition, assume Condition (C5) is violated. Then there should exist a vector $\gamma$ with unit length $||\gamma||=1$. In the meanwhile, we have $\gamma^{\top}(\mB^{\top}\mB)\gamma=\lambda_{\max}(\mB^{\top}\mB)\to\infty$ as $p\to \infty$. Then we have $\var(Z_i^{\top}\gamma)=\var(X_i^{\top}\mB\gamma)=\gamma^\top \mB^{\top}E(X_iX_i^{\top})\mathbb B \gamma\geq C_{\min}\{\gamma^{\top}(\mB^{\top}\mB)\gamma\}=C_{\min}\lambda_{\max}(\mB^{\top}\mB) \to\infty$. In this case, the information contained in $Z_i$ as measured by $\var(Z_i^{\top}\gamma)$ diverges to infinity. Thus we can make nearly perfect prediction for $Y_i$ by $Z_i$. This is obviously not reasonable. Therefore, we must have Condition (C5) hold. With the above conditions, we then have the following theorem.

\begin{theorem}
	Assume the conditions (C1)-(C5), then $\|\widehat{\mathbb{B}}_{\text{mle}}^{\text{Source}} - \mathbb{B}\|_F = O_p(\sqrt{p/N})$.\label{Theorem 1}
\end{theorem}
\noindent
The technical proof of Theorem \ref{Theorem 1} is similar to that of \cite{fan2004nonconcave}. However, for the sake of theoretical completeness, we provide the detailed proof in Appendix A. By Theorem \ref{Theorem 1}, we know the maximum likelihood estimator $\widehat{\mathbb{B}}_{\text{mle}}^{\text{Source}}$ of the source model is $\sqrt{N/p}$-consistent. It implies that, with a huge source sample size, the parameter $\mathbb{B}$ can be estimated with reasonable accuracy. Once $\widehat{\mathbb{B}}_{\text{mle}}^{\text{Source}}$ is computed, a lower-dimensional feature vector can be computed as $\widehat Z_i = \widehat{\mathbb{B}}_{\text{mle}}^{\text{Source}\top}X_i$. Then a revised target model with $\widehat Z_i$ as the input can be established. The theoretical properties of the corresponding estimator $\widehat \gamma$ are summarized in the following theorem.

\begin{theorem}
	Assume the conditions (C1)-(C5), then we have: (1) $\sqrt n(\widehat \gamma - \gamma) \stackrel{d}{\to } N(0,I(\gamma)^{-1})$; (2) $\sqrt n v^\top\Big(\wh \theta^{\text{Transfer}} - \theta\Big)/\sigma(v) \stackrel{d}{\to} N(0,1)$, where $v$ is an arbitrary $p$-dimensional vector in $\mS(\mathbb{B})$ with unit length, and $\sigma^2(v)=v^\top\mathbb B I(\gamma)^{-1}\mathbb B^\top v$. \label{Theorem 2}
\end{theorem}
\noindent
The proof of Theorem \ref{Theorem 2} can be found in Appendix B. Theorem \ref{Theorem 2} suggests that, with the help of the source model estimator $\widehat{\mathbb{B}}_{\text{mle}}^{\text{Source}}$, the estimator $\widehat \gamma$ of the target model should be $\sqrt n$-consistent and asymptotically normal. In particular, its asymptotic distribution remains to be the same as that of the oracle estimator, which is defined as $\widehat{\gamma}_{\text{oracle}}=\argmax_{\gamma} {\mL}_n(\gamma; Z)$. This is the maximum likelihood estimator obtained with the true low dimensional feature vector $Z_i$ instead of its estimator $\widehat{Z}_i$.
Furthermore, with the relationship $\theta = \mathbb{B} \gamma$, the TL estimator of $\theta$ can be obtained as  $\widehat{\theta}^{\text{Transfer}} = \widehat{\mathbb{B}}_{\text{mle}}^{\text{Source}} \widehat \gamma$.
Then Theorem \ref{Theorem 2} suggests that, for any $v \in \mS(\mathbb{B})$, the estimator $v^{\top}\widehat{\theta}^{\text{Transfer}}$ is $\sqrt n$-consistent for $v^{\top}\theta$ and asymptotically normal.

\csubsection{Testing Sufficiency for Transfer Learning}

The nice theoretical properties obtained in the previous section hinge on one critical assumption. That is, all information contained in $X_i$ about $Y_i$ in the target data can be sufficiently represented by $Z_i = \mathbb{B}^\top X_i$. In other words, the transfer learning is sufficient. However, for a practical dataset, whether the transfer learning is sufficient is not immediately clear. Thus there is great need to develop a principled statistical procedure to test for the transfer learning sufficiency. We are then motivated to develop in this section a novel solution in this regard.

Our method is motivated by the following interesting observation. If the transfer learning is sufficient in the sense that all information contained in $X_i$ about $Y_i$ has been sufficiently captured by $Z_i$, we should have $E(Y_i|Z_i,X_i) = E(Y_i|Z_i) = g(Z_i^\top\gamma)$. Then the pseudo residual $\ve_i=Y_i - E(Y_i |Z_i, X_i)=Y_i-g(Z_i^\top\gamma)$ should be uncorrelated with the feature vector $X_i$ in the sense that $E(\ve_iX_i) = 0$. Practically, the true feature vector $Z_i$ and the coefficient parameter $\gamma$ are both unknown. However, with the help of source data, they can be both estimated as $\widehat Z_i = \widehat{\mathbb{B}}_{\text{mle}}^{\text{Source}\top}X_i$ and $\widehat\gamma = \argmax_{\gamma} {\mL}_n(\gamma;\widehat Z)$. Then the estimated pseudo residual can be obtained as $\widehat{\ve}_i=Y_i-g(\widehat Z_i^\top\widehat{\gamma})$. We then should reasonably expect $n^{-1}\sum_{i=1}^n\widehat{\ve}_iX_i\approx 0$. This leads to a preliminary test statistic as $T_1 = \|n^{-1} \sum_{i=1}^{n} \widehat\ve_i X_i\|^2=n^{-2} \widehat\ve^\top \mathbb X\mathbb X^\top \widehat\ve$, where $\widehat\ve = (\widehat \ve_1,\dots,\widehat\ve_n)^\top\in\mR^n$ is the estimated pseudo residual vector and $\mathbb X = (X_1,\dots,X_n)^\top \in \mR^{n\times p}$ is the design matrix. If the null hypothesis of transfer learning sufficiency is incorrect, we should expect $\widehat\ve_i$ to be materially correlated with $X_{ij}$ for some $1\leq j \leq p$. This should make the value of $T_1$ unreasonably large. Similar test statistics were also constructed by \cite{lan2014testing} but for testing statistical significance of a high dimensional regression model. To understand the asymptotic behavior of $T_1$, it is important to compute its mean and variance under the null hypothesis of transfer
learning sufficiency. This leads to the following theorem.
\begin{theorem}
	Assume the conditions (C1)-(C4) and the null hypothesis of transfer learning sufficiency. Define $\Sigma_\gamma = E(\ve_i^2 X_iX_i^\top) = E[g(Z_i^\top\gamma)\{1-g(Z_i^\top\gamma)\}X_iX_i^\top ]$. We further assume that $n^2 p/ N \to 0$. Then we have $T_1=T^*_1+o_p(\sqrt{p}/n)$, where $T_1^*=n^{-2}\ve^\top \mX\mX^\top\ve$ and $\ve = (\ve_1,\dots,\ve_n)^\top\in\mR^n$ is the true pseudo residual vector. For $T^*_1$, we have $E(T^*_1) = n^{-1}\tr(\Sigma_\gamma)$ and $\var(T^*_1) = n^{-3}\var(\ve_i^2 X_i^\top X_i) + 2n^{-2}(1-n^{-1})\tr(\Sigma_\gamma^2)$. \label{Theorem 3}
\end{theorem}

\noindent
The proof of Theorem \ref{Theorem 3} can be found in Appendix C. By Theorem \ref{Theorem 3} and Condition (C3), we have $E(T_1^*)\ge (2n)^{-1}\tr(\Sigma_\gamma) \ge 0.5C_{\min} n^{-1}p$. Further define $\text{SE}(T_1^*)=\{\var(T_1^*)\}^{1/2}$. Then by Theorem \ref{Theorem 3} and Condition (C3), we can obtain $\text{SE}(T_1^*) \le \big\{4n^{-3} E(X_iX_i^\top)^4 + 4n^{-2}(1-n^{-1}) \tr(\Sigma_\gamma^2)\big\}^{1/2} \le 2c_{\max}(n^{-3/2}p +n^{-1}p^{1/2})$, where $c_{\max} = \max\{C_{\max}, \tau_{\max}\}$. Therefore, we should have $E(T_1^*) /\text{SE}(T_1^*) \to \infty$ as $n\to\infty$. Consequently, it is impossible for the random variable $T_1 / \text{SE}(T_1) $ to converge in distribution to any non-degenerated probability distribution. Similar interesting phenomenon was also obtained in \cite{lan2014testing}. To solve this problem, applying appropriate centralization on $T_1$ is needed.

\csubsection{A Centralized Test Statistic}

As discussed above, the order of $E(T_1)$ is considerably larger than that of $\text{SE}(T_1)$. Therefore, appropriate centralization on $T_1$ is inevitably needed. Similar operations have been used in the past literature \citep{li2007nonparametric,lan2014testing}. Note that the leading term of $E(T_1)$ is linearly related to $tr(\Sigma_\gamma)$ where $\Sigma_\gamma = E(\ve_i^2 X_i X_i^\top)$. Therefore, the bias of $T_1$ can be significantly reduced by subtracting this term, as long as an accurate estimator for $\Sigma_\gamma$ can be obtained. In this regard, a simple moment estimator of $\Sigma_\gamma$ can be constructed as $\widehat \Sigma_\gamma = n^{-1} \sum_{i=1}^{n}\widehat\ve_i^2 X_i X_i^\top$. This leads to a centralized test statistic as $T_2 = T_1 - n^{-1}\tr(\widehat \Sigma_\gamma)$. The mean and variance of $T_2$ are then given by the following theorem.

\begin{theorem}
Assume the conditions (C1)-(C4) and the null hypothesis of transfer learning sufficiency. Further assume $n^2 p/ N \to 0$. Then we have $T_2=T_2^*+o_p(\sqrt{p}/n)$, where $T_2^*=n^{-2}\ve^\top (\mX\mX^\top - \mathbb D_X)\ve$ and $\mathbb D_X = \diag\{X_1^\top X_1,\dots,X_n^\top X_n\}\in \mR^{n\times n}$. For $T_2^*$, we have $E(T_2^*) = 0$ and $\var(T_2^*) = 2n^{-2}\tr(\Sigma_\gamma^2)\{1+o(1)\}$. \label{Theorem 4}
\end{theorem}

\noindent
The detailed proof of Theorem \ref{Theorem 4} can be found in Appendix D. By Theorem \ref{Theorem 4}, we obtain the following interesting findings. First, we find that the asymptotical variance of the centralized statistic $T_2^*$ is significantly smaller than that of $T_1^*$, by an amount of $n^{-3}\var(\ve_i^2 X_i^\top X_i)=O_p(p^2/n^3)$. Second, note that
$\text{SE}(T_2^*)=\{\var(T_2^*)\}^{1/2}= \{2n^{-2}\tr(\Sigma_\gamma^2)\}^{1/2}\ge C_{\min}n^{-1}p^{1/2}$. Compared with
the order of $\text{SE}(T_2^*)$, we find the order of $E(T_2^*)$ is a small order term. Therefore a formal standardized test statistic can be constructed as $T_3 = T_2 / \sqrt{2n^{-2}\tr(\Sigma_\gamma^2)}$, whose asymptotic behavior is summarized in the following theorem.

\begin{theorem}
	Assume the conditions (C1)-(C4) and the null hypothesis of transfer learning sufficiency. Further assume $n^2 p/ N \to 0$, then $T_3 \stackrel{d}{\to} N(0,1)$ as $N\to \infty$.\label{Theorem 5}
\end{theorem}
\noindent
The detailed proof of Theorem \ref{Theorem 5} can be found in Appendix E. Theorem \ref{Theorem 5} suggests that, the centralized and standardized test statistic $T_3$ is asymptotically distributed as a standard normal distribution. Note that the term $\tr(\Sigma_\gamma^2)$ is involved in $T_3$. Therefore, to practically apply this test procedure, a ratio consistent estimator for $\tr(\Sigma_\gamma^2)$ is needed. To solve this problem, one natural idea is to use the plug-in estimator $\tr(\widehat\Sigma_\gamma^2)$. The expectation of this plug-in estimator can be calculated as $E\big\{\tr(\widehat\Sigma_\gamma^2)\big\} = n^{-1}E\big\{\ve_i^4(X_i^\top X_i)^2\} + (1-n^{-1})\tr(\Sigma_\gamma^2)\{1+o(1)\big\}$.
By Condition (C3), we can derive $n^{-1}E\big\{\ve_i^4(X_i^\top X_i)^2\big\} \ge (2n)^{-1}\big\{E(\ve_i^2 X_i^\top X_i)\big\}^2 = (2n)^{-1}\tr^2(\Sigma_\gamma) \ge C_{\min}^2(2n)^{-1}p^2$. We can also have $\tr(\Sigma_\gamma^2) \le C_{\max}^2p$. This suggests that $E\{\tr(\widehat{\Sigma}_\gamma^2)\}/\tr(\Sigma_\gamma^2)\geq p/n$. As a result, when $p$ is larger than $n$, $\tr(\widehat \Sigma_\gamma^2)$ cannot be ratio consistent for $\tr(\Sigma_\gamma^2)$ in the sense that $\tr(\widehat \Sigma_\gamma^2)/\tr(\Sigma_\gamma^2) \stackrel{p}{\nrightarrow} 1$ as $N {\to}\infty$. The inconsistency of $\tr(\widehat \Sigma_\gamma^2)$ is mainly due to the extra bias term $n^{-1}E\big\{\ve_i^4(X_i^\top X_i)^2\big\}$. Therefore, we are motivated to construct a bias corrected estimator for $\tr(\Sigma_\gamma^2)$ as
\[
\widehat{\tr(\Sigma_\gamma^2)} = \tr(\widehat \Sigma_\gamma^2) - n^{-2}\sum_{i=1}^{n}\widehat\ve_i^4\big(X_i^\top X_i\big)^2 = n^{-2}\sum_{i\ne j}\widehat\ve_i^2\widehat\ve_j^2 \big(X_i^\top X_j\big)^2.
\]
The ratio consistency property of $\widehat{\tr(\Sigma_\gamma^2)}$ is summarized by the following theorem.
\begin{theorem}
	Assume the conditions (C1)-(C4) and the null hypothesis of transfer learning sufficiency. Further assume $n^2 p/ N \to 0$, then $\widehat{\tr(\Sigma_\gamma^2)}/\tr(\Sigma_\gamma^2)\stackrel{p}{\to}1$  as $N {\to} \infty$.\label{Theorem 6}
\end{theorem}
\noindent
The detailed proof of Theorem \ref{Theorem 6} can be found in Appendix F. Theorem \ref{Theorem 6} implies that $\widehat{\tr(\Sigma_\gamma^2)}$ is a ratio consistent estimator for $\tr(\Sigma_\gamma^2)$ as $N {\to} \infty$. With this ratio consistent estimator, we can construct a practically computable test statistic as ${T}_4 = T_2 / \{2n^{-2}\widehat{\tr}(\Sigma_\gamma^2)\}^{1/2}$. Under the null hypothesis of transfer
learning sufficiency, we should have $T_4$ to follow a standard normal distribution asymptotically. Then the null hypothesis of transfer learning sufficiency should be rejected, if the value of $T_4$ is sufficiently large. Specifically, we should reject the null hypothesis at a given significance level $\alpha$ if ${T}_4 > z_{1-\alpha}$, where $z_\alpha$ is the $\alpha$th quantile of the standard normal distribution. Otherwise, the null hypothesis of transfer learning sufficiency should be accepted.

\csection{NUMERICAL STUDIES}

\csubsection{Simulation Setups}

To demonstrate the finite sample performance of the TL estimator and the test statistic for transfer learning sufficiency, we present here a number of simulation studies. The whole study contains two parts. The first part focuses on the estimation performance of the TL estimator $\widehat\theta^{\text{Transfer}}$. The objective is to numerically investigate the influence of feature dimension $p$, target sample size $n$, and the source sample size $N$ on the statistical efficiency of the TL estimator under the null hypothesis of transfer learning sufficiency. The second part investigates the finite sample performance of the proposed test statistic $T_4$ for transfer learning sufficiency. Both the size (i.e., the Type I error) and the power are evaluated under different experimental settings of $(p,n,N)$. For the entire study, we fix the class number $K=8$ in the source data. Then with a given feature dimension $p$, target sample size $n$, and source sample size $N$, the detailed data generation procedure is present as follows.

{\sc Source Data.} We first generate the covariate $X_i^* \in \mR^{p}$ with $1\leq i \leq N$ independently and identically from a multivariate normal distribution with mean 0 and covariance matrix $\Sigma = (\sigma_{j_1j_2})_{p\times p}$ with $\sigma_{j_1j_2} = 0.5^{|j_1-j_2|}$.
Then we generate a random vector $\tilde\beta_k \in \mR^p$ for $0\leq k \leq K$. All the elements of $\tilde\beta_k$ are independently and identically generated from a standard normal distribution. Next we define $\beta_k = \tilde{\beta}_k/\|\tilde{\beta}_k\| - \tilde\beta_0/\|\tilde\beta_0\|$. Thereafter the source response $Y_i^*$ is generated according to $P(Y_i^* = k|X_i^*) = \exp(X_i^{*\top}\beta_k) / \big\{1+\sum_{j=1}^{K} \exp(X_i^{*\top}\beta_j)\big\}$ for $1\le k\le K$. Accordingly, we have $P(Y_i^* = 0|X_i^*) = 1 / \big\{1+\sum_{j=1}^{K} \exp(X_i^{*\top}\beta_j)\big\}$. Let $\mathbb B = (\beta_1,\dots,\beta_k)\in \mR^{p\times K}$. We then have a feature generating function as $Z_{\mathbb B}(X_i^*)=\mathbb B^\top X_i^*$.

{\sc Target Data.}
We first generate the target covariate $X_i \in \mR^{p}$ with $1\leq i \leq n$ independently and identically from the same multivariate normal distribution as for the source data.
Next we compute the low dimensional feature vector $Z_i =Z_{\mathbb B}(X_i)= \mathbb B^\top X_i \in \mR^{K}$ for the target data. Then we consider how to generate the response $Y_i$. Specifically, under the null hypothesis of transfer learning sufficiency, we generate $Y_i$ according to $P(Y_i =1 |X_i)=P(Y_i =1 |Z_i) = \exp(Z_i^\top\gamma)/\{1+\exp(Z_i^\top\gamma)\}$ with $\gamma = (0.5,0.5,0.5,0.5,0.5,-1.25,0,0)^\top \in \mR^{K}$. Therefore, the true coefficient $\theta$ associated with $X_i$ in the target mode is given by $\theta = \mathbb{B} \gamma$. In contrast, if the transfer learning is insufficient, we then generate $Y_i$ as $P(Y_i = 1|X_i) = \exp(Z_i^\top\gamma + X_{i1}\delta) / \{1+\exp(Z_i^\top\gamma + X_{i1}\delta) \}$, where $\delta > 0$ is a fixed constant.

\csubsection{Results of Estimation Performance}

\begin{figure}[H]
	\centering
	\subfloat[$n=600$ and $N = 40000$]{\includegraphics[width=0.33\textwidth]{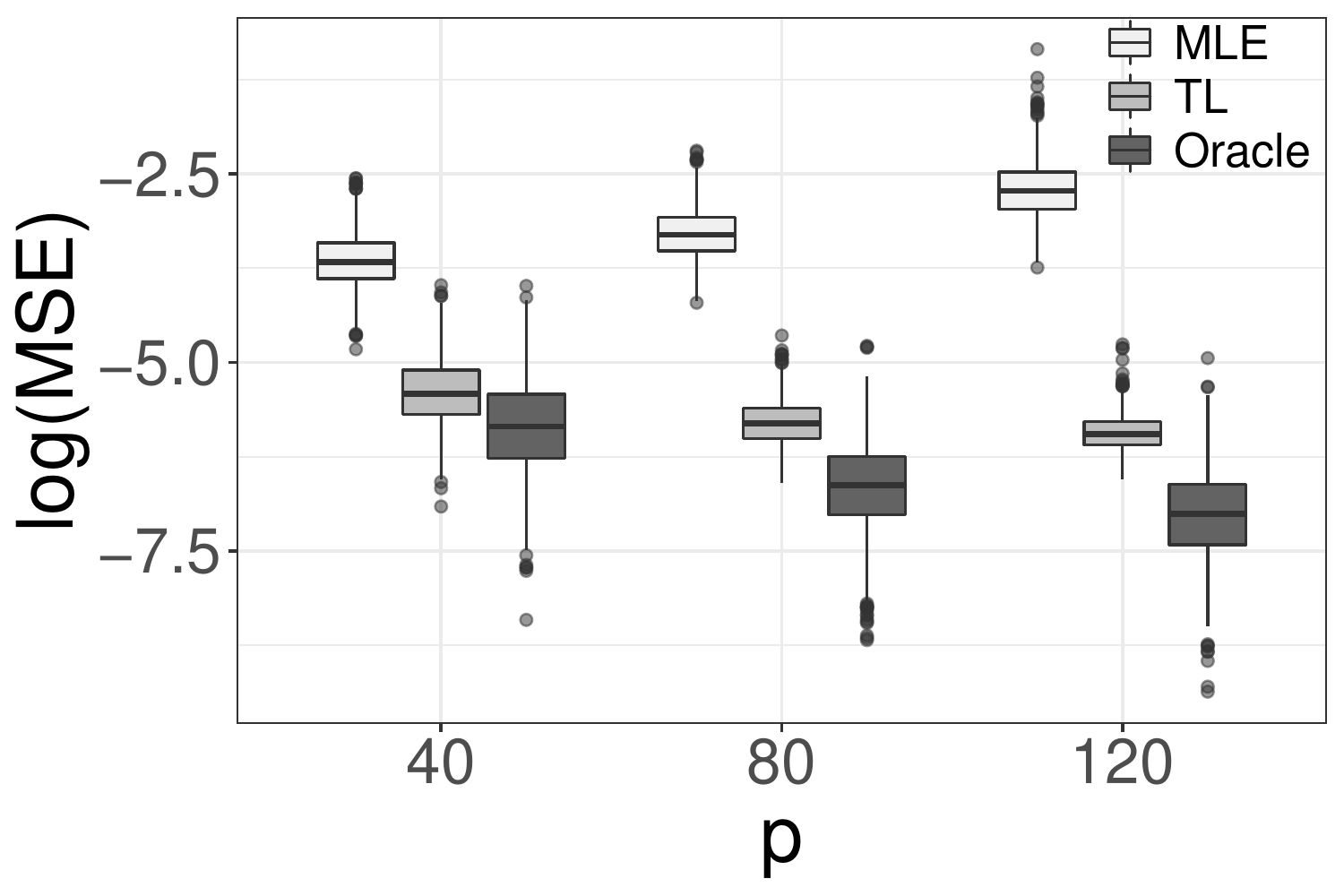}}
	\subfloat[$p = 80$ and $N = 40000$]{\includegraphics[width=0.33\textwidth]{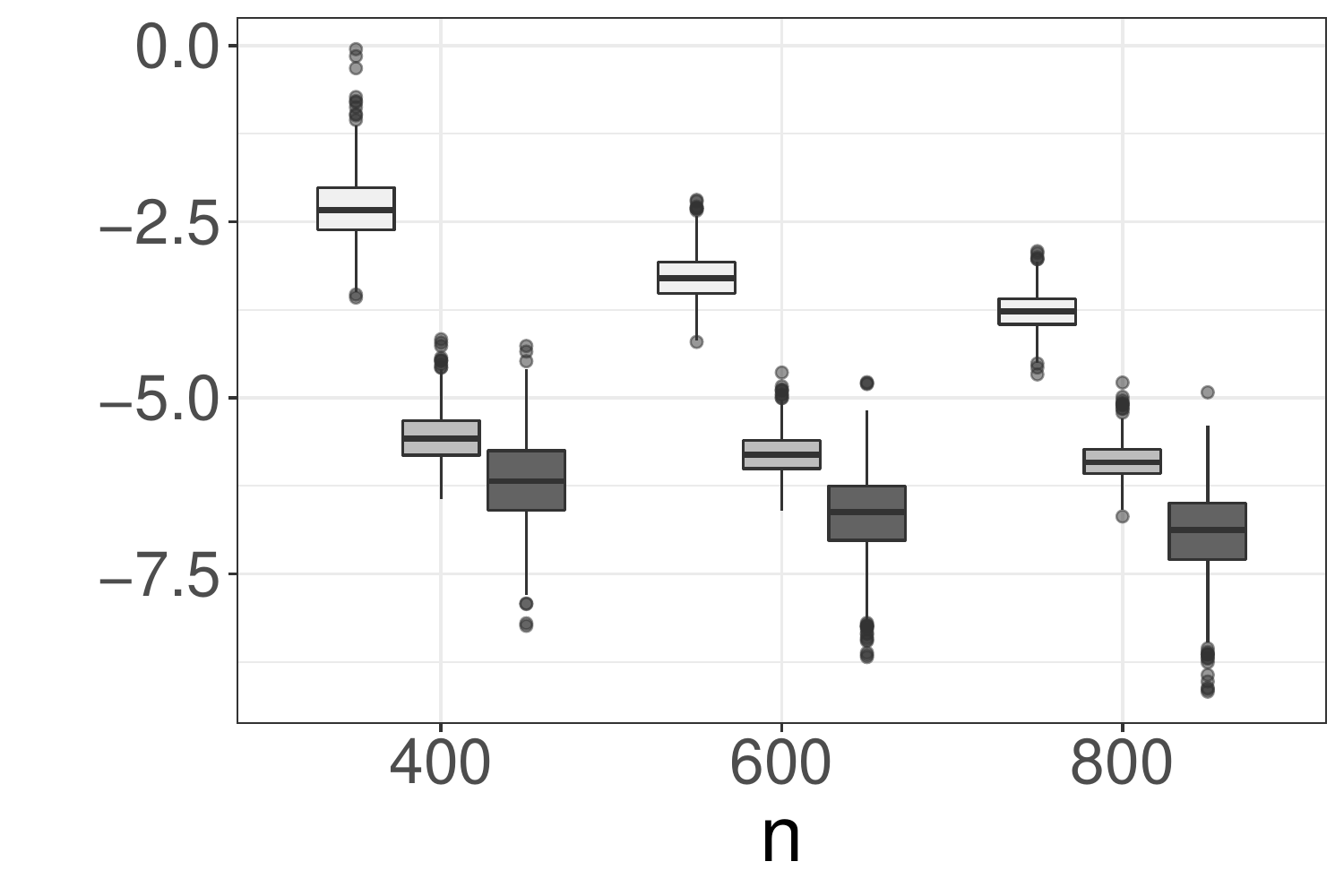}}
	\subfloat[$p = 80$ and $n = 600$]{\includegraphics[width=0.33\textwidth]{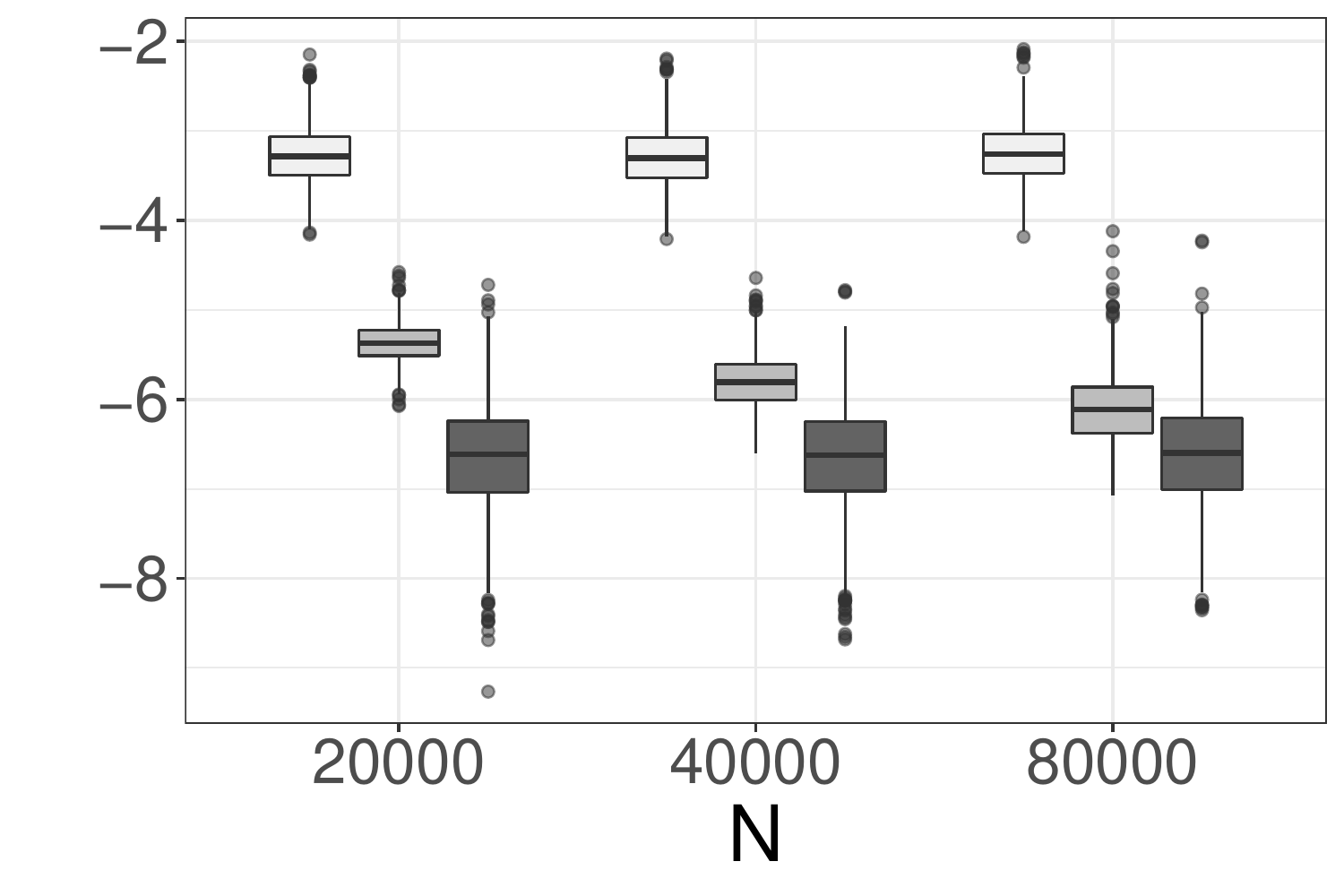}}
	\caption{Boxplots of log-transformed MSE of the maximum likelihood estimator (MLE), our proposed transfer learning (TL) estimator, and the oracle transfer learning estimator in the $p<n$ case. Different combinations of $(p,n,N)$ are considered to explore the influence of $(p,n,N)$ on the estimation performance of the three estimators.}
	\label{Figure 1}
\end{figure}

We investigate the finite sample performance of the proposed TL estimator under the null hypothesis of transfer learning sufficiency. We start with the case $p<n$. In this case, we compare three different estimators of $\theta$. They are, respectively: (1) our proposed TL estimator $\widehat\theta^{\text{Transfer}}$; (2) the oracle estimator $\widehat\theta^{\text{Oracle}}$, which is also a transfer learning estimator but using the true coefficient matrix $\mathbb B$ instead of its estimated counterpart $\widehat{\mathbb{B}}_{\text{mle}}^{\text{Source}} $; and (3) the target data based maximum likelihood estimator $\widehat{\theta}_{\text{mle}}^{\text{Target}}$ according to \eqref{target likelihood}.
Let the feature dimension $p \in (40,80,120)$, the target sample size $n\in (400,600,800)$, and the source sample size $N\in (20000,40000,80000)$. Then different combinations of $(p,n,N)$ are studied. For each $(p,n,N)$ combination, we generate the target data and source data according to the generation process as described in Section 3.1. The experiment is randomly replicated for a total of $B=1,000$ times. For the $b$-th replication, denote one particular estimator of $\theta$ as $\widehat\theta^{(b)}$. Then we evaluate its estimation performance by mean squared error (MSE) as $\text{MSE}_{(b)}=p^{-1}\sum_{j=1}^p(\widehat\theta^{(b)}_j-\theta_j)^2$. This leads to a total of $B=1,000$ MSE values. They are then log-transformed and boxplotted in Figure \ref{Figure 1}.

By Figure \ref{Figure 1}, we can draw the following conclusions. First, for all cases, the log(MSE) values of the MLE are always larger than those of our proposed TL estimator and the oracle estimator. This is expected, because the MLE is obtained by directly estimating a high dimensional model without any help from the source data. As a consequence, its performance is worse than the two transfer learning estimators. Second, with the feature dimension $p$ and source sample size $N$ fixed, the log(MSE) values of the TL estimator, along with the other estimators, decrease as the target sample size $n$ increases. These results demonstrate the consistency of the TL estimator, which is in line with Theorem \ref{Theorem 2}.
Last, with the feature dimension $p$ and target sample size $n$ fixed, we find the log(MSE) values of the TL estimator become closer to those of the oracle estimator as $N\to \infty$. This result also corroborates our theoretical findings in Theorem \ref{Theorem 2}. That is, when the source sample size is large enough, the TL estimator can perform as well as the oracle estimator asymptotically.

\begin{figure}[h]
	\centering
	\subfloat[$n=100$ and $N = 40000$]{\includegraphics[width=0.33\textwidth]{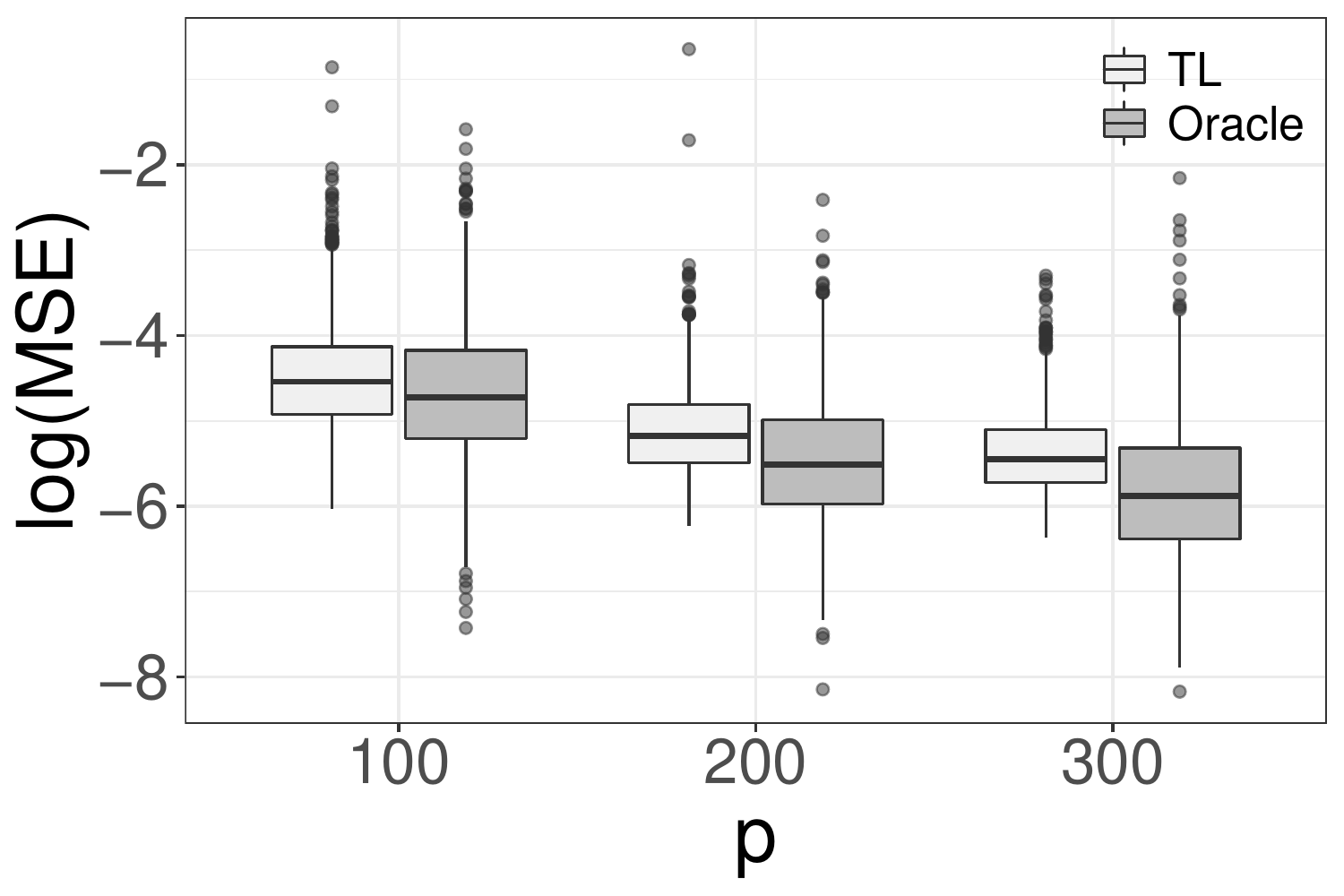}}
	\subfloat[$p = 300$ and $N = 40000$]{\includegraphics[width=0.33\textwidth]{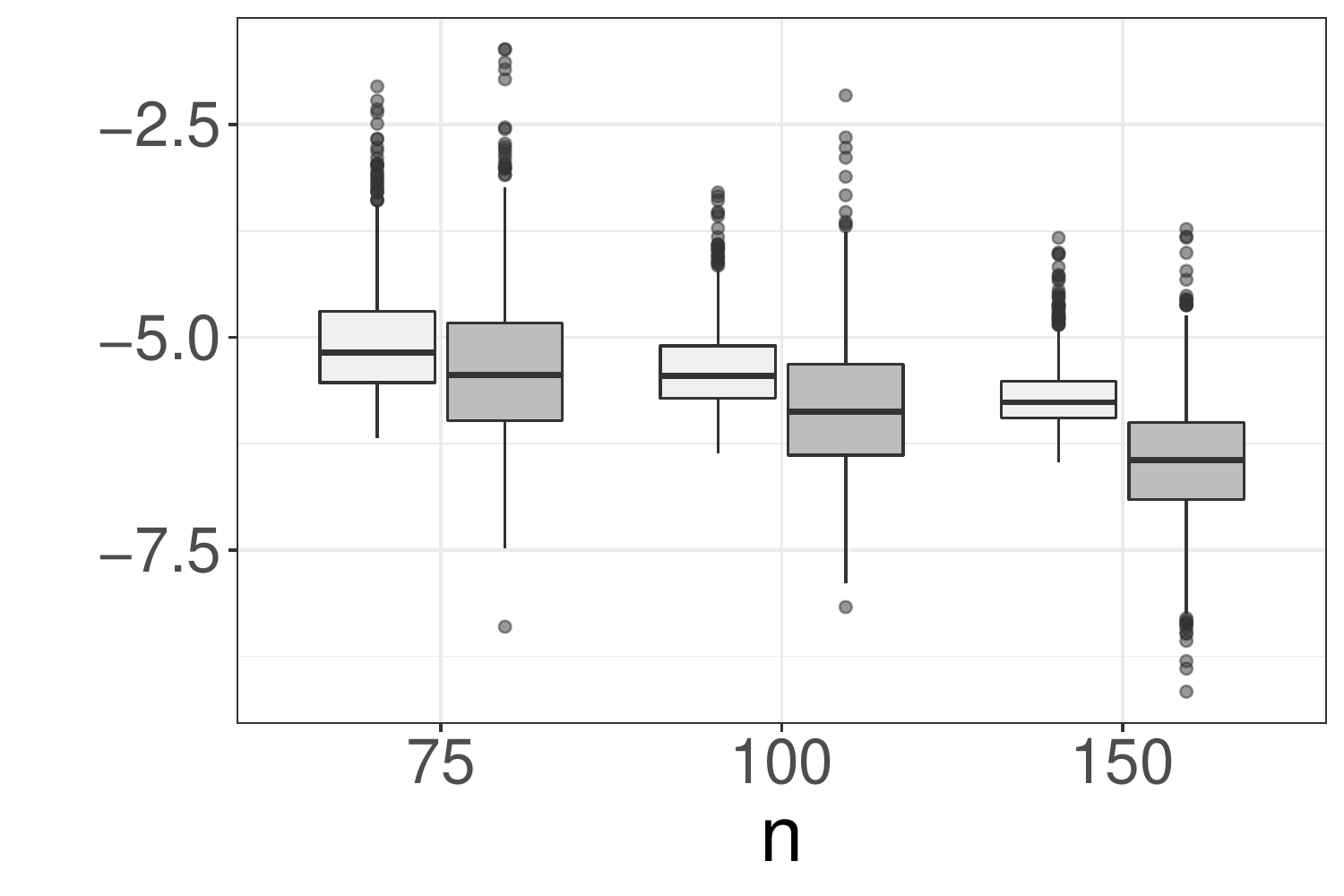}}
	\subfloat[$p = 300$ and $n = 100$]{\includegraphics[width=0.33\textwidth]{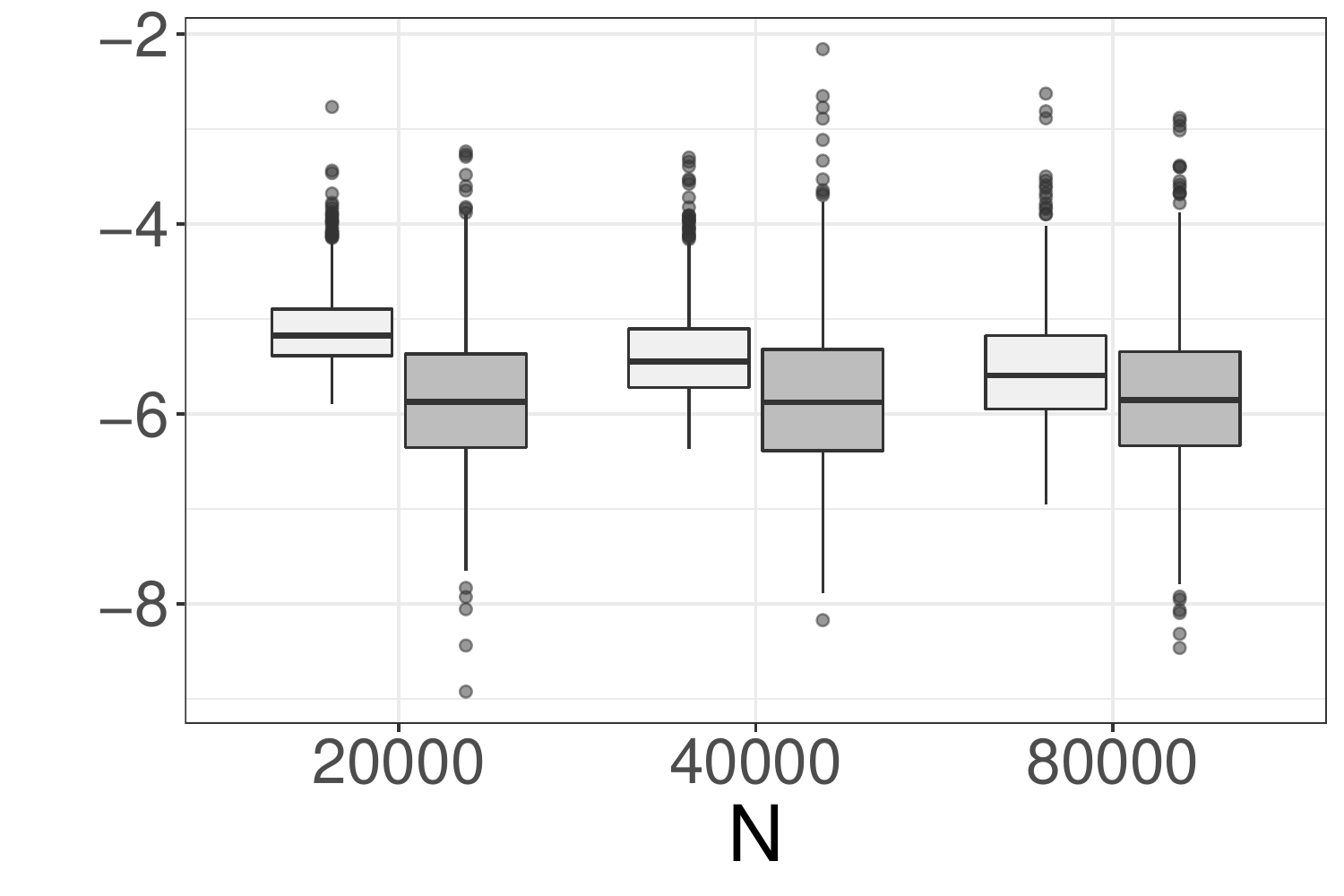}}
	\caption{Boxplots of log-transformed estimation error of our proposed transfer learning (TL) estimator and the oracle transfer learning estimator in the $n\le p < N$ case. Different combinations of $(p,n,N)$ are considered to explore the influence of $(p,n,N)$ on the estimation performance of the two estimators.}
	\label{Figure 2}
\end{figure}

Next consider the case with $n\le p < N$. In this case, the MLE $\widehat{\theta}_{\text{mle}}^{\text{Target} }$ is not computable. Therefore, we only compare the estimation performance of our proposed TL estimator $\widehat{\theta}^{\text{Transfer}}$ with the oracle estimator $\widehat\theta^{\text{Oracle}}$. We consider the feature dimension $p\in (100,200,300)$, the target sample size $n \in (75,100,150)$, and the source sample size $N \in (20000,40000,80000)$. Then different combinations of $(p,n,N)$ are studied; see Figure \ref{Figure 2} for details. For each combination, we randomly replicate the experiment for a total of $B=1,000$ times. For each experiment, the two transfer learning estimators are computed and the MSE values are calculated to evaluate the estimation performance. This leads to $B=1,000$ MSE values for each estimator, which are then log-transformed and boxplotted in Figure \ref{Figure 2}. Similar with the case $p<n$, we find the log(MSE) values of the two estimators both decrease as the target sample size $n$ increases, implying the consistency the two estimators. In addition, the two estimators become closer to each other as $N\to \infty$. This finding suggests that, once we provide the source data with a sufficiently large sample size, the high dimensionality effect on the estimation performance of $\theta$ is very limited.

\csubsection{Results of Testing Performance}

In this section, we investigate the finite sample performance of the proposed testing procedure for transfer learning sufficiency. We first evaluate the size of the testing procedure. In this regard, we should generate the target data under the null hypothesis of transfer learning sufficiency. Specifically, we generate $Y_i$ according to $P(Y_i =1 |X_i) =P(Y_i =1 |Z_i) = \exp(Z_i^\top\gamma)/\{1+\exp(Z_i^\top\gamma)\}$. In this case, we compare our proposed test statistic with the oracle one, which is computed using the true $Z_i$ instead of $\wh Z_i$. We consider the feature dimension $p \in (1000, 1500,2000,3000)$, the target sample size $n \in (200,300,500)$, and the source sample size $N \in (400000,800000)$. For each $(p,n,N)$ combination, we randomly replicate the experiment for a total of $B = 1,000$ times. Let $ T_4^{(b)}$ be the proposed test statistic computed in the $b$th random replication. We then compute the empirical rejection probability as $\mathrm{EJP} = B^{-1} \sum_{b=1}^B I(T_4^{(b)} > z_{1-\alpha})$. Then $\mathrm{EJP}$ corresponds to the empirical size (i.e., Type I error), since the target data are generated under the null hypothesis. The EJP of the oracle test statistic can be computed similarly. Table \ref{Table 1} presents the empirical sizes of our proposed test statistic and the oracle test statistic under the significance level $\alpha = 0.05$. We find the empirical sizes of the two test statistics are fairly close to the nominal null $5\%$, as long as the source sample is sufficiently large. These results confirm our theoretical claims in Section 2.4.

\begin{table}[h]
	\caption{The empirical sizes of our proposed transfer learning sufficiency test, as well as the oracle test under the significance level $\alpha = 0.05$}\label{Table 1}
\small
\renewcommand\arraystretch{1.3}
	\centering
	\begin{tabular}{cc|cccc|cccc}
		\hline
		\hline
		&                    & \multicolumn{4}{c|}{Oracle Test} & \multicolumn{4}{c}{Empirical Test} \\ \hline
		$n$   &  \diagbox{$N$}{$p$} & 1000   & 1500   & 2000   & 3000  & 1000   & 1500   & 2000   & 3000   \\ \hline
		\multirow{3}{*}{200}& $4\times 10^5$             & 0.025  & 0.037  & 0.053  & 0.036 & 0.033  & 0.047  & 0.053  & 0.039  \\
		& $6\times 10^5$      & 0.030  & 0.032  & 0.038  & 0.031
		& 0.037  & 0.048  & 0.045 & 0.032  \\
		& $8\times 10^5$             & 0.034  & 0.039  & 0.047  & 0.038 & 0.033  & 0.038  & 0.041  & 0.039  \\ \hline
		\multirow{3}{*}{300}& $4\times 10^5$             & 0.037  & 0.043  & 0.040  & 0.040  & 0.053  & 0.044  & 0.048  & 0.058  \\
		& $6\times 10^5$  & 0.032 & 0.045 & 0.042 & 0.040
		 &  0.036 & 0.049  & 0.044  & 0.053 \\
		& $8\times 10^5$             & 0.022  & 0.039  & 0.041   & 0.039 & 0.028  & 0.040   & 0.047  & 0.045  \\ \hline
		\multirow{3}{*}{500}& $4\times 10^5$             & 0.033  & 0.038  & 0.033  & 0.032 & 0.053  & 0.048  & 0.057  & 0.052  \\
		& $6\times 10^5$         & 0.039 & 0.037  & 0.045 & 0.047
		& 0.049  & 0.049  & 0.055  & 0.067   \\
		& $8\times 10^5$              & 0.026  & 0.041  & 0.033  & 0.030  & 0.033  & 0.048  & 0.042  & 0.041  \\ \hline
	\end{tabular}
\end{table}

\begin{figure}[H]
	\centering
	\includegraphics[width=0.7\textwidth]{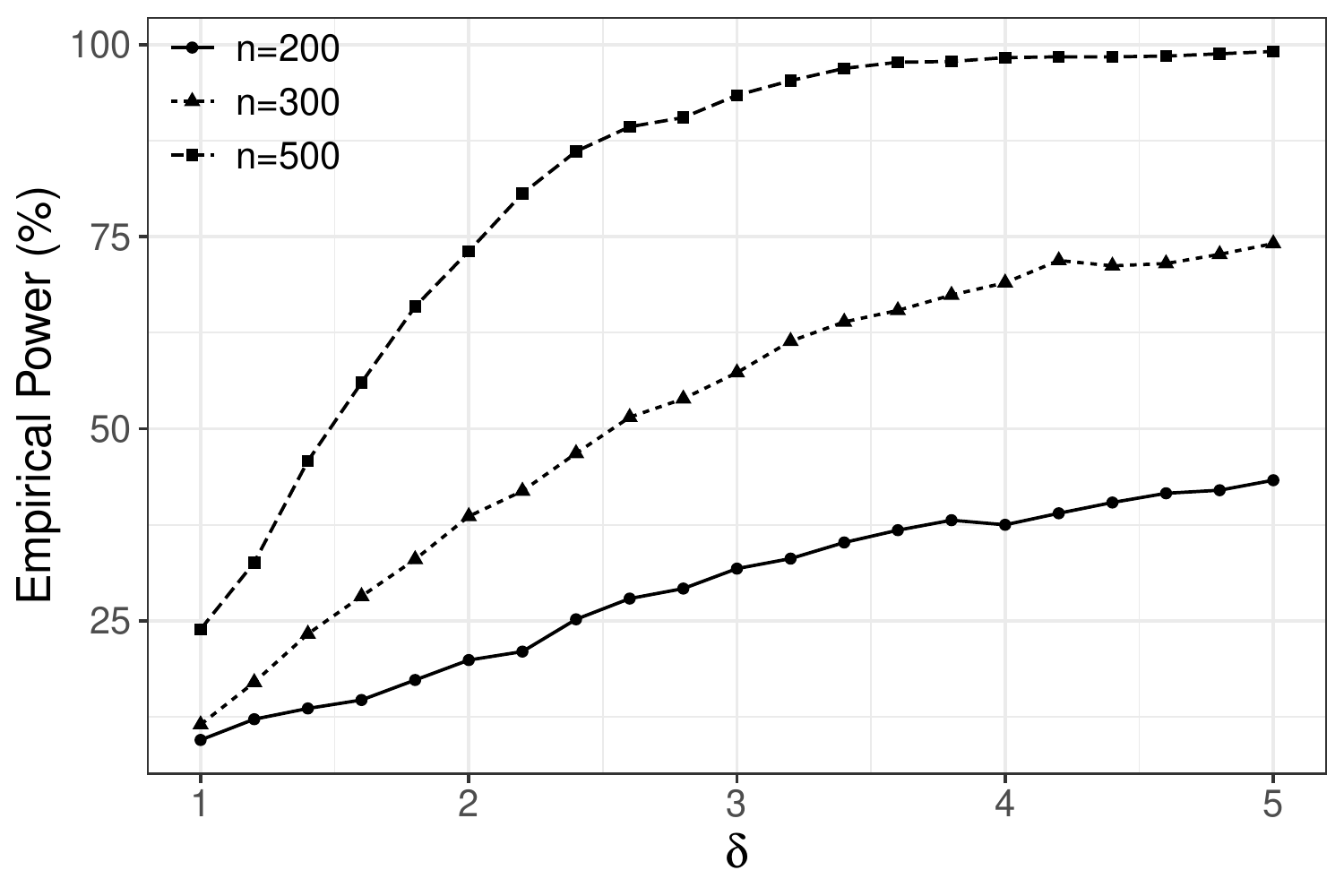}
	\caption{Empirical power of our proposed transfer learning sufficiency test with different target sample size $n$ and signal strength $\delta$. Here we set $p = 2000$, $N=400000$, and the significance level $\alpha = 0.05$.}
	\label{Figure 3}
\end{figure}

Next, we evaluate the power of the proposed testing procedure for transfer learning sufficiency. Therefore, we should generate the target data under the alternative hypothesis, i.e., the transfer learning insufficiency. In this case, we generate $Y_i$ according to $P(Y_i = 1|X_i) = \exp(Z_i^\top\gamma + X_{i1}\delta) / \{1+\exp(Z_i^\top\gamma + X_{i1}\delta)\}$ with $\delta> 0$. We refer to $\delta$ as the signal strength. A larger $\delta$ value indicates the larger effect of $X_{i1}$ on $Y_i$. Thus there exists more information left in $X_i$, which cannot be captured by $Z_i$ and therefore leads to stronger evidence to support the alternative hypothesis. We fix $p=2,000$ and $N=400,000$. Three target sample sizes are considered as $n \in  (200,300,500)$. We also let $\delta$ vary from 1 to 5 with the step size 0.2. For each combination of $(n,\delta)$, we repeat the experiment for a total of $B=1,000$ times. For each experiment, we compute the proposed test statistics $T_4$ and then calculate the empirical rejection probability $\mathrm{EJP}$ in the $B$ replications under the significance level $\alpha = 0.05$. The $\mathrm{EJP}$ corresponds to the empirical power of the hypothesis testing procedure, since the alternative hypothesis holds. Figure \ref{Figure 3} displays the empirical power of the proposed test statistic in different experimental settings. By Figure \ref{Figure 3}, we find larger signal strength $\delta$ leads to better empirical power. In addition, larger target sample size $n$ leads to larger empirical power.

\csubsection{A Real Data Example}

To further demonstrate the proposed testing method, we present here a real data example. The target data used here is the \textit{Dogs vs.Cats} dataset, which is downloadable in the Kaggle competition (\url{https://www.kaggle.com/competitions/dogs-vs-cats/}). This dataset contains $n = 25,000$ images belonging to two categories (i.e., dogs or cats). The goal here is to classify each image into the two categories, which is denoted by a binary response $Y_i$. This is a standard image classification problem, which can be addressed by various deep learning models \citep{NIPS2012c399862d,he2016deep,Howard2017MobileNetsEC}. To demonstrate our method, we first convert the unstructured image data into structured vectors. To this end, various pretrained deep learning models are used. Specifically, we take the feature vectors generated by the last second layer of the pretrained models as our covariate $X_i$, whose dimension is usually high. Given the fact that the target sample size is limited as compared with the feature dimension, we turn to transfer learning for help.

We consider the \textit{ImageNet} dataset \citep{2014ImageNet} as our source data. This is a dataset containing a total of $N = 1,281,167$ images and covering a total of $K+1 = 1000$ categories. To convert the images in the source data into structured vectors, the same pretrained deep learning models as for the target data are used. This leads to the high dimensional source covariate $X_i^*$. Thereafter, a standard multi-class logistic regression model can be fitted for $X_i^*$ and $Y_i^*$, where $Y_i^* \in \{0, 1,..., K\}$ indicating the class label for the $i$th image in the source data. This leads to the source data based maximum likelihood estimator $\wh{\mathbb{B}}_{\text{mle}}^{\text{Source}}$. To transfer information from the source data, a low dimensional feature vector $\wh Z_i = \wh{\mathbb{B}}_{\text{mle}}^{\text{Source}\top} X_i \in \mR^K$ can be constructed for the target data. Thereafter, a standard binary logistic regression model is fitted for $\wh Z_i$ and $Y_i$. Accordingly, the proposed testing procedure for transfer learning sufficiency can be applied and the prediction accuracy can be evaluated.

For the sake of comprehensive evaluation, a number of classical pretrained deep learning models are used to extract the high dimensional feature vectors $X_i$ and $X_i^*$. They are, respectively, the VGG \citep{simonyan2014very} model with two different sizes, the ResNet \citep{he2016deep} model with three different sizes, and the InceptionV3 \citep{szegedy2016rethinking}. The resulting feature dimensions are given by $p = 4,096$ for the VGG and $p = 2,048$ for both the ResNet and InceptionV3. The original target dataset has been randomly split into two parts, among which one is used for training with the sample size 15,000 and the other one is used for testing with the sample size 10,000. The proposed testing procedure for transfer learning sufficiency is then applied on the training dataset. The resulting $p$-values are then reported in Table \ref{Table 2} for every pretrained deep learning model. The prediction accuracy is also evaluated but on the testing dataset. The detailed results are reported in Table \ref{Table 2}. By Table \ref{Table 2}, we find that the $p$-values of all tests are larger than the significance level $\alpha = 0.05$. These results support the null hypostypsis of transfer learning sufficiency. This seems to be very reasonable testing results since the prediction accuracies evaluated on the test dataset are all very high.

\begin{table}[htbp]
	\caption{The $p$-values of the proposed testing procedure and the prediction accuracies on the testing dataset are reported under different pretrained deep learning models.}
	\centering
\begin{tabular}{cccc}
	\hline
	\hline
	Model & $p$  & $p$-Value & Prediction Accuracy\\ \hline
	VGG16 & 4096   &0.846 &97.97\%\\
	VGG19 & 4096  & 0.809 & 97.93\%\\
	\hline
	ResNet50 & 2048  & 0.827 & 98.06\%\\
	ResNet101 & 2048     &0.777   & 98.40\%   \\
	ResNet152 & 2048   &0.755 &98.50\%\\
	\hline
	InceptionV3 & 2048  &0.691 & 98.97\%\\
	\hline
\end{tabular}
	\label{Table 2}
\end{table}

\csection{CONCLUDING REMARKS}

We present here a principled methodology for testing sufficiency of transfer learning. In this regard, a transfer learning estimation procedure is developed and a testing statistic is constructed, whose asymptotic null distribution is analytically derived. Extensive numerical studies are presented to demonstrate the finite sample performance of the proposed method. To conclude this article, we consider here some interesting topics for future study. First, the testing procedure studied in this work requires the sample size of the source data to be much larger than that of the target data. However, in real practice, the sample size of the source data might be comparable with that of the target data. Then how to test for transfer learning sufficiency in this case needs further investigation. Second, if the null hypothesis of transfer learning sufficiency is rejected, how to further improve the estimation accuracy of the target model becomes the next important problem. Last, we use linear transformation in the feature generating function, which might be too restrictive. Then how to test for transfer learning sufficiency with nonlinear feature generating functions is another important problem worthwhile further studying.

\renewcommand\refname{\begin{center}
		\large{REFERENCES}
\end{center}}
\bibliographystyle{asa}
\bibliography{reference}

\begin{thebibliography}{40}
\newcommand{\enquote}[1]{``#1''}
\expandafter\ifx\csname natexlab\endcsname\relax\def\natexlab#1{#1}\fi

\bibitem[{Cai and Wei(2021)}]{Cai2021Transfer}
Cai, T.~T. and Wei, H. (2021), \enquote{Transfer Learning for Nonparametric
  Classification: Minimax Rate and Adaptive Classifier,} \textit{The Annals of
  Statistics}, 49, 100--128.

\bibitem[{Candes and Sur(2020)}]{EJ2020THE}
Candes, E.~J. and Sur, P. (2020), \enquote{The Phase Transition for the
  Existence of the Maximum Likelihood Estimate in High-dimensional Logistic
  Regression,} \textit{The Annals of Statistics}, 48, 27--42.

\bibitem[{Donahue et~al.(2014)Donahue, Jia, Vinyals, Hoffman, Zhang, Tzeng, and
  Darrell}]{donahue2014decaf}
Donahue, J., Jia, Y., Vinyals, O., Hoffman, J., Zhang, N., Tzeng, E., and
  Darrell, T. (2014), \enquote{Decaf: A Deep Convolutional Activation Feature
  for Generic Visual Recognition,} in \textit{International conference on
  machine learning}, PMLR, pp. 647--655.

\bibitem[{Fan and Fine(2013)}]{Fan2013Linear}
Fan, C. and Fine, J.~P. (2013), \enquote{Linear Transformation Model with
  Parametric Covariate Transformations,} \textit{Journal of the American
  Statistical Association}, 108, 701--712.

\bibitem[{Fan and Li(2001)}]{fan2001variable}
Fan, J. and Li, R. (2001), \enquote{Variable selection via nonconcave penalized
  likelihood and its oracle properties,} \textit{Journal of the American
  statistical Association}, 96, 1348--1360.

\bibitem[{Fan and Peng(2004)}]{fan2004nonconcave}
Fan, J. and Peng, H. (2004), \enquote{Nonconcave penalized likelihood with a
  diverging number of parameters,} \textit{The Annals of Statistics}, 32,
  928--961.

\bibitem[{Farahani et~al.(2020)Farahani, Pourshojae, Rasheed, and
  Arabnia}]{Farahani2020A}
Farahani, A., Pourshojae, B., Rasheed, K., and Arabnia, H.~R. (2020),
  \enquote{A Concise Review of Transfer Learning,} in \textit{2020
  International Conference on Computational Science and Computational
  Intelligence (CSCI)}, pp. 344--351.

\bibitem[{Guo et~al.(2018)Guo, Lei, Xing, Yan, and Li}]{guo2018deep}
Guo, L., Lei, Y., Xing, S., Yan, T., and Li, N. (2018), \enquote{Deep
  convolutional transfer learning network: A new method for intelligent fault
  diagnosis of machines with unlabeled data,} \textit{IEEE Transactions on
  Industrial Electronics}, 66, 7316--7325.

\bibitem[{Hall and Heyde(1980)}]{hall1980martingale}
Hall, P. and Heyde, C.~C. (1980), \textit{Martingale limit theory and its
  application}, Academic press.

\bibitem[{He et~al.(2016)He, Zhang, Ren, and Sun}]{he2016deep}
He, K., Zhang, X., Ren, S., and Sun, J. (2016), \enquote{Deep Residual Learning
  for Image Recognition,} in \textit{Proceedings of the IEEE conference on
  computer vision and pattern recognition}, pp. 770--778.

\bibitem[{He and Shao(2000)}]{he2000parameters}
He, X. and Shao, Q.-M. (2000), \enquote{On parameters of increasing
  dimensions,} \textit{Journal of Multivariate Analysis}, 73, 120--135.

\bibitem[{Howard et~al.(2017)Howard, Zhu, Chen, Kalenichenko, Wang, Weyand,
  Andreetto, and Adam}]{Howard2017MobileNetsEC}
Howard, A.~G., Zhu, M., Chen, B., Kalenichenko, D., Wang, W., Weyand, T.,
  Andreetto, M., and Adam, H. (2017), \enquote{MobileNets: Efficient
  Convolutional Neural Networks for Mobile Vision Applications,}
  \textit{ArXiv}, abs/1704.04861.

\bibitem[{Ji et~al.(2021)Ji, Guo, Yang, Zhang, and Lu}]{2021Multi}
Ji, J., Guo, Y., Yang, Z., Zhang, T., and Lu, X. (2021), \enquote{Multi-level
  Dictionary Learning for Fine-grained Images Categorization with Attention
  Model,} \textit{Neurocomputing}, 453, 403--412.

\bibitem[{Krizhevsky et~al.(2012)Krizhevsky, Sutskever, and
  Hinton}]{NIPS2012c399862d}
Krizhevsky, A., Sutskever, I., and Hinton, G.~E. (2012), \enquote{ImageNet
  Classification with Deep Convolutional Neural Networks,} in \textit{Advances
  in Neural Information Processing Systems}, eds. Pereira, F., Burges, C.,
  Bottou, L., and Weinberger, K., Curran Associates, Inc., vol.~25.

\bibitem[{Lan et~al.(2014)Lan, Wang, and Tsai}]{lan2014testing}
Lan, W., Wang, H., and Tsai, C.~L. (2014), \enquote{Testing Covariates in
  High-dimensional Regression,} \textit{Annals of the Institute of Statistical
  Mathematics}, 66, 279--301.

\bibitem[{Li and Racine(2007)}]{li2007nonparametric}
Li, Q. and Racine, J.~S. (2007), \textit{Nonparametric econometrics: theory and
  practice}, Princeton University Press.

\bibitem[{Li et~al.(2022{\natexlab{a}})Li, Cai, and Li}]{li2020transfer}
Li, S., Cai, T.~T., and Li, H. (2022{\natexlab{a}}), \enquote{Transfer Learning
  for High-dimensional Linear Regression: Prediction, Estimation, and Minimax
  Optimality,} \textit{Journal of the Royal Statistical Society, Series B}, 84,
  149--173.

\bibitem[{Li et~al.(2022{\natexlab{b}})Li, Cai, and Li}]{Li2022Transfer}
--- (2022{\natexlab{b}}), \enquote{Transfer Learning in Large-scale Gaussian
  Graphical Models with False Discovery Rate Control,} \textit{Journal of the
  American Statistical Association}, in the press.

\bibitem[{Liang et~al.(2022)Liang, Cai, Sun, and Xia}]{Liang2022locally}
Liang, Z., Cai, T.~T., Sun, W., and Xia, Y. (2022), \enquote{Locally Adaptive
  Transfer Learning Algorithms for Large-Scale Multiple Testing,} \textit{arXiv
  preprint:2203.11461}.

\bibitem[{Lu and Zhang(2010)}]{Lu2010On}
Lu, W. and Zhang, H.~H. (2010), \enquote{On Estimation of Partially Linear
  Transformation Models,} \textit{Journal of the American Statistical
  Association}, 105, 683--691.

\bibitem[{McCullagh and Nelder(1989)}]{McCullagh1989}
McCullagh, P. and Nelder, J. (1989), \textit{Generalized Linear Models, 2nd
  Edition}, Chapman \& Hall/CRC.

\bibitem[{Oquab et~al.(2014)Oquab, Bottou, Laptev, and Sivic}]{6909618}
Oquab, M., Bottou, L., Laptev, I., and Sivic, J. (2014), \enquote{Learning and
  Transferring Mid-level Image Representations Using Convolutional Neural
  Networks,} in \textit{Proceedings of the 2014 IEEE Conference on Computer
  Vision and Pattern Recognition}, pp. 1717--1724.

\bibitem[{Pan and Yang(2009)}]{pan2009survey}
Pan, S.~J. and Yang, Q. (2009), \enquote{A Survey on Transfer Learning,}
  \textit{IEEE Transactions on Knowledge and Data Engineering}, 22, 1345--1359.

\bibitem[{Portnoy(1985)}]{portnoy1985asymptotic}
Portnoy, S. (1985), \enquote{Asymptotic behavior of $ M $ estimators of $ p $
  regression parameters when $ p^{2} /n $ is large; II. Normal approximation,}
  \textit{The Annals of Statistics}, 13, 1403--1417.

\bibitem[{Russakovsky et~al.(2014)Russakovsky, Deng, Su, Krause, Satheesh, Ma,
  Huang, and et~al.}]{2014ImageNet}
Russakovsky, O., Deng, J., Su, H., Krause, J., Satheesh, S., Ma, S., Huang, Z.,
  and et~al. (2014), \enquote{ImageNet large scale visual recognition
  challenge,} in \textit{CoRR, abs/1409.0575}.

\bibitem[{Seber(2008)}]{seber2008matrix}
Seber, G.~A. (2008), \textit{A matrix handbook for statisticians}, John Wiley
  \& Sons.

\bibitem[{Shao(2003)}]{shao2003mathematical}
Shao, J. (2003), \textit{Mathematical statistics}, Springer Science \& Business
  Media.

\bibitem[{Shin et~al.(2016)Shin, Roth, Gao, Lu, Xu, Nogues, Yao, Mollura, and
  Summers}]{shin2016deep}
Shin, H.-C., Roth, H.~R., Gao, M., Lu, L., Xu, Z., Nogues, I., Yao, J.,
  Mollura, D., and Summers, R.~M. (2016), \enquote{Deep convolutional neural
  networks for computer-aided detection: CNN architectures, dataset
  characteristics and transfer learning,} \textit{IEEE transactions on medical
  imaging}, 35, 1285--1298.

\bibitem[{Simonyan and Zisserman(2014)}]{simonyan2014very}
Simonyan, K. and Zisserman, A. (2014), \enquote{Very deep convolutional
  networks for large-scale image recognition,} \textit{arXiv preprint
  arXiv:1409.1556}.

\bibitem[{Sur and Candes(2019)}]{Sur2019A}
Sur, P. and Candes, E.~J. (2019), \enquote{A modern maximum-likelihood theory
  for high-dimensional logistic regression,} \textit{Proceedings of the
  National Academy of Sciences}, 119, 14516--14525.

\bibitem[{Szegedy et~al.(2016)Szegedy, Vanhoucke, Ioffe, Shlens, and
  Wojna}]{szegedy2016rethinking}
Szegedy, C., Vanhoucke, V., Ioffe, S., Shlens, J., and Wojna, Z. (2016),
  \enquote{Rethinking the inception architecture for computer vision,} in
  \textit{Proceedings of the IEEE conference on computer vision and pattern
  recognition}, pp. 2818--2826.

\bibitem[{Tian and Feng(2022)}]{Tian2022Transfer}
Tian, Y. and Feng, Y. (2022), \enquote{Transfer Learning under High-dimensional
  Generalized Linear Models,} \textit{Journal of the American Statistical
  Association}, 1, 1--14.

\bibitem[{Van~der Vaart(1998)}]{vaart_1998}
Van~der Vaart, A.~W. (1998), \textit{Asymptotic Statistics}, Cambridge Series
  in Statistical and Probabilistic Mathematics, Cambridge University Press.

\bibitem[{Vershynin(2018)}]{vershynin2018high}
Vershynin, R. (2018), \textit{High-dimensional probability: An introduction
  with applications in data science}, vol.~47, Cambridge university press.

\bibitem[{Wainwright(2019)}]{wainwright2019high}
Wainwright, M.~J. (2019), \textit{High-dimensional statistics: A non-asymptotic
  viewpoint}, vol.~48, Cambridge University Press.

\bibitem[{Wang(2011)}]{wang2011gee}
Wang, L. (2011), \enquote{GEE analysis of clustered binary data with diverging
  number of covariates,} \textit{The Annals of Statistics}, 39, 389--417.

\bibitem[{Wang et~al.(2021)Wang, Yao, Kwok, and Ni}]{10.1145/3386252}
Wang, Y., Yao, Q., Kwok, J.~T., and Ni, L.~M. (2021), \enquote{Generalizing
  from a Few Examples: A Survey on Few-Shot Learning,} \textit{ACM Computing
  Surveys}, 53, 1--34.

\bibitem[{Weiss et~al.(2016)Weiss, Khoshgoftaar, and Wang}]{Weiss2016A}
Weiss, K., Khoshgoftaar, T.~M., and Wang, D. (2016), \enquote{A Survey of
  Transfer Learning,} \textit{Journal of Big Data}, 3, 1--34.

\bibitem[{Welsh(1989)}]{welsh1989m}
Welsh, A. (1989), \enquote{On M-processes and M-estimation,} \textit{The Annals
  of Statistics}, 337--361.

\bibitem[{Zhong and Chen(2011)}]{zhong2011tests}
Zhong, P.-S. and Chen, S.~X. (2011), \enquote{Tests for high-dimensional
  regression coefficients with factorial designs,} \textit{Journal of the
  American Statistical Association}, 106, 260--274.

\end{thebibliography}
\newpage
\appendix
\renewcommand{\csection}[1]
{\begin{center}
		\stepcounter{section}
		{\bf\large Appendix \Alph{section}. #1}
	\end{center}
}

\scsection{APPENDIX}

\csection{Proof of Theorem \ref{Theorem 1}}

Let $\mathbb{U} = (u_1,\dots,u_K) \in \mR^{p\times K}$ be a matrix with unit length in the sense that $\|\mathbb U\|_F = 1$, where $u_k = (u_{k1},\dots,u_{kp})^\top\in\mR^p$ is the $k$th column vector of $\mathbb{U}$. Next, by \cite{fan2001variable}, it suffices to show that for an arbitrary $\ve >0$, we have
\beq
\liminf_{n\to\infty}P\Big\{\sup_{\|\mathbb{U}\|_F = 1} \mathcal L_N^{\text{Source}}\big(\mathbb{B}+ C \sqrt{p/N}\mathbb{U}) < \mathcal L_N^{\text{Source}}\big(\mathbb{B})\Big\} \ge 1-\varepsilon \label{liminf of vecB}
\eeq
for a sufficiently large but fixed constant $C>0$. By Taylor expansion, we have
\begin{gather}
p^{-1}\Big\{\mL_N^{\text{Source}}\Big(\mathbb{B}+ C \sqrt{p/N}\mathbb{U}\Big) - \mL_N^{\text{Source}}\big(\mathbb{B}\big)\Big\}
= C(Np)^{-1/2}\sum_{k=1}^K \bigg(\frac{\partial \mL_N^{\text{Source}}(\mathbb{B})}{\partial\beta_k}\bigg)^\top u_k\nonumber \\ + 2^{-1}C^2 N^{-1}\sum_{k=1}^K\sum_{l=1}^K u_k^\top \Big(\frac{\partial^2 \mL_N^\text{Source}(\mathbb{B})}{\partial \beta_k \partial \beta_l^\top}\Big) u_l \nonumber \\
 + 6^{-1}C^3p^{1/2}N^{-3/2} \sum_{j=1}^K\frac{\partial }{\partial\beta_j}\Big\{\sum_{k=1}^K\sum_{l=1}^K u_k^\top \Big(\frac{\partial^2 \mL_N^\text{Source}(\mathbb{\bar B})}{\partial \beta_k \partial \beta_l^\top}\Big) u_l\Big\}^\top u_j \label{vecB Taylor},
\end{gather}
where $\bar{\mathbb{B}} = \mathbb{B}+ \nu C \sqrt{p/N}\mathbb{U}$ for some $\nu \in (0,1)$. We next study the three terms in \eqref{vecB Taylor} separately. By the Cauchy-Schwarz inequality, we have $|\sum_{k=1}^K(\partial \mL_N^{\text{Source}}(\mathbb{B})/\partial \beta_k)^\top u_k| \le \{\sum_{k=1}^K \|u_k\|^2\}^{1/2}\{\sum_{k=1}^K \|\partial \mL_N^{\text{Source}}(\mathbb{B})/\partial \beta_k\|^2\}^{1/2}  = \{\sum_{k=1}^K \|\partial \mL_N^{\text{Source}}(\mathbb{B})/\partial \beta_k\|^2\}^{1/2}$. Direct computation leads to $E\|\partial \mL_N^{\text{Source}}(\mathbb{B})/\partial \beta_k\|^2 = E(\sum_{i=1}^N \pi_{ik}(1-\pi_{ik}) X_i^{*\top}X_i^* ) \le N E(X_i^{*\top}X_i^* )= N E\{\tr(X_i^{*\top}X_i^*)\}= N\tr\{E(X_i^*X_i^{*\top})\} \le Np\lambda_{\max}\{E(X_i^*X_i^{*\top})\} = O(Np)$ by Condition (C3). This leads to $\|\partial \mL_N^{\text{Source}}(\mathbb{B})/\partial \beta_k\|  = O_p(\sqrt{Np})$ . As a consequence, we have $\sqrt{1/Np}\sum_{k=1}^K(\partial \mL_N^{\text{Source}}(\mathbb{B})/\partial \beta_k)^\top u_k = O_p(1)$. Then we consider the second term of \eqref{vecB Taylor}. For any $\eta >0$ and $k\in\{1,\dots,K\}$, by Chebyshev's inequality, we have
\begin{gather*}
	P\bigg[\Big\|N^{-1}\sum_{i=1}^N \pi_{ik} (1-\pi_{ik}) X_i^* X_i^{*\top } - E\Big\{\pi_{ik}(1-\pi_{ik}) X_i^*X_i^{*\top} \Big\}\Big\|_F > \eta \bigg]
	\\ \le\eta^{-2} \sum_{j=1}^p \sum_{l=1}^p E\bigg[N^{-1}\sum_{i=1}^N \pi_{ik} (1-\pi_{ik}) X_{ij}^*X_{il}^* - E\Big\{\pi_{ik} (1-\pi_{ik}) X_{ij}^* X_{il}^* \Big\} \bigg]^2 = O(p^2/N).
\end{gather*}
By Condition (C1), we know that $p^2/N \le C^2n^2/N \to 0$. As a result, we have $\|N^{-1}\sum_{i=1}^N \pi_{ik} (1-\pi_{ik}) X_i^* X_i^{*\top } - E\{\pi_{ik}(1-\pi_{ik}) X_i^*X_i^{*\top}\}\|_F = o_p(1)$.
Similarly, for $k,l\in \{1,\dots,K\}$ and $k\neq l$, we can show that $\|N^{-1}\sum_{i=1}^N \pi_{ik} \pi_{il} X_i^* X_i^{*\top } - E(\pi_{ik} \pi_{il} X_i^*X_i^{*\top} )\|_F = o_p(1)$.
Recall that $W_i = \diag(\pi_i) - \pi_i\pi_i^\top \in \mR^{K\times K}$ with $\pi_i = (\pi_{i1},\dots,\pi_{iK})^\top$ and $\pi_{ik} = P(Y_i^* = k|X_i^*)$. Combining the above results, we have
\begin{equation}
\begin{split}
&\lambda_{\min}^2\bigg[N^{-1}\sum_{i=1}^NW_i \otimes \big(X_i^*X_i^{*\top}\big) -  E\Big\{W_i\otimes \big(X_i^*X_i^{*\top}\big)\Big\}\bigg]\\
	\le &\Big\|N^{-1}\sum_{i=1}^NW_i \otimes \big(X_i^*X_i^{*\top}\big) - E\Big\{W_i\otimes \big(X_i^*X_i^{*\top}\big)\Big\}\Big\|_F^2\\
	= &\sum_{k=1}^K\Big\|N^{-1}\sum_{i=1}^N \pi_{ik} (1-\pi_{ik}) X_i^* X_i^{*\top } - E\Big\{\pi_{ik}(1-\pi_{ik}) X_i^*X_i^{*\top} \Big\}\Big\|_F^2\\
	+&\sum_{k\neq l}\Big\|N^{-1}\sum_{i=1}^N \pi_{ik} \pi_{il} X_i^* X_i^{*\top } - E\Big(\pi_{ik} \pi_{il} X_i^*X_i^{*\top} \Big)\Big\|_F^2 = o_p(1)\nonumber
\end{split}
\end{equation}
By Weyl's inequality \citep{seber2008matrix}, we know that $\lambda_{\min}(A+B)\ge \lambda_{\min}(A) + \lambda_{\min}(B)$ for any two symmetric matrices $A$ and $B$. As a result, for the second term of \eqref{vecB Taylor}, we have
\begin{equation}
\begin{split}
\label{min eigen}
&-N^{-1}\sum_{k=1}^K\sum_{l=1}^K u_k^\top \frac{\partial^2 \mL_N^\text{Source}(\mathbb{B})}{\partial \beta_k \partial \beta_l^\top} u_l -\lambda_{\min}\bigg[E\Big\{W_i\otimes \big(X_i^*X_i^{*\top}\big)\Big\}\bigg] \\
 = &\mathrm{vec}(\mathbb{U})^\top \Big\{N^{-1}\sum_{i=1}^NW_i \otimes \big(X_i^*X_i^{*\top}\big) \Big\} \mathrm{vec}(\mathbb{U}) - \lambda_{\min}\bigg[E\Big\{W_i\otimes \big(X_i^*X_i^{*\top}\big)\Big\}\bigg] \\
\ge &\lambda_{\min}\Big\{N^{-1}\sum_{i=1}^NW_i \otimes \big(X_i^*X_i^{*\top}\big)\Big\} - \lambda_{\min}\bigg[E\Big\{W_i\otimes \big(X_i^*X_i^{*\top}\big)\Big\}\bigg] \\
\ge &\lambda_{\min}\bigg[N^{-1}\sum_{i=1}^NW_i \otimes \big(X_i^*X_i^{*\top}\big) -  E\Big\{W_i\otimes \big(X_i^*X_i^{*\top}\big)\Big\}\bigg] \stackrel{p}{\to} 0.
\end{split}
\end{equation}

By the Condition (C3), we know that $E\{W_i\otimes (X_i^*X_i^{*\top})\}$ is a positive definite matrix with $\lambda_{\min}\{E(W_i\otimes X_i^*X_i^{*\top})\} \ge C_{\min}$ for some positive constant $C_{\min}>0$. As a consequence,  we have $-N^{-1}\sum_{k=1}^K\sum_{l=1}^K u_k^\top \partial^2 \mL_N^\text{Source}(\mathbb{B})/\partial \beta_k \partial \beta_l^\top u_l > 0$ with probability tending to one. For the third term of \eqref{vecB Taylor}, by Cauchy-Schwarz inequality,
\begin{gather*}
\Big|\frac{\partial }{\partial\beta_j}\Big\{\sum_{k=1}^K\sum_{l=1}^K u_k^\top \Big(\frac{\partial^2 \mL_N^\text{Source}(\mathbb{\bar B})}{\partial \beta_k \partial \beta_l^\top}\Big) u_l\Big\}^\top u_j \Big|
= \Big| \sum_{k=1}^K\sum_{l=1}^K \sum_{i=1}^N\Big(\frac{\partial p_{ikl}}{\partial \beta_j}  \Big)^\top u_j\Big(X_i^{*\top}u_k\Big)\Big(X_i^{*\top}u_l\Big) \Big|\\
\le \sum_{k=1}^K\sum_{l=1}^K  \sup_{i}\Big\|\frac{\partial p_{ikl}}{\partial \beta_j}\Big|_{\beta_j = \bar\beta_j} \Big\| \Big\{\sum_{i=1}^N(u_k^\top X_i^*X_i^{*\top } u_k)\Big\}^{1/2}\Big\{\sum_{i=1}^N(u_l^\top X_i^*X_i^{*\top} u_l)\Big\}^{1/2},
\end{gather*}
where $p_{ikl} = \pi_{ik}(1-\pi_{ik})$ if $k=l$ and $p_{ikl} = -\pi_{ik}\pi_{il}$ if $k\neq l$. By simple computation, in the case of $k=l$, we have $\partial p_{ikl} / \partial \beta_j =  \pi_{ik}(1-\pi_{ik})(1-2\pi_{ik})X_i^*$ if $j=k=l$, and $\partial p_{ikl} / \partial \beta_j = -\pi_{ik}\pi_{ij}(1-2\pi_{ik})X_i^*$ if $j\neq k$. In the case of $k\neq l$, we have $\partial p_{ikl} / \partial \beta_j = 2\pi_{ik}\pi_{il}\pi_{ij}X_i^*$ if $j\neq k$, and $j\neq l$ and $\partial p_{ikl} / \partial \beta_j = -\pi_{ik}\pi_{il}(1-2\pi_{ij})X_i^*$ if $j= k$ or $j= l$. In all cases, we have $\sup_i\|\partial p_{ikl} / \partial \beta_j \|\le 2\sup_i\|X_i^{*}\|$. Note that by Condition (C2), we know that each component of $X_i^*$ is sub-Gaussian. Thus, by Proposition 2.5.2 in \cite{vershynin2018high}, we know that $E\exp(X_{ij}^{*2}/\|X^*_{ij}\|_{\psi_2}^2) \le 2$. Let $j^*_i = \max\{j: X_{ij}^{*2} /\|X^*_{ij} \|_{\psi_2}^2\}$. Then by the fact that $\|X_i^*\|_{\psi_2} \ge \|X^*_{ij}\|_{\psi_2}$, we have 
$E\exp\{\|X_i^*\|^2/(p\|X_i^*\|_{\psi_2}^2)\} \le E\exp (X_{ij^*}^{*2} / \|X^*_{ij^*}\|_{\psi_2}^2) \le 2$. Then we know that $\|X_i^*\| / \sqrt p$ is sub-Gaussian. By the results of Exercise 2.5.10 in \cite{vershynin2018high}, we then have $\sup_i\|X_i^*\| = O_p(\sqrt{p\log N})$. As a result, we have $\sup_i\|\partial p_{ikl} / \partial \beta_j \| = O_p(\sqrt{p\log N})$. By a similar proof of \eqref{min eigen}, we know that $\lambda_{\max}(N^{-1}\sum_{i=1}^NX_i^*X_i^{*\top}) - \lambda_{\max}\{E(X_i^*X_i^{*\top})\}\stackrel{p}{\to}0$. By Condition (C3), we know that $\lambda_{\max}\{E(X_i^*X_i^{*\top})\} \le C_{\max}$. Note that $\|u_k\|\le 1$. Then we have $N^{-1}\sum_{i=1}^N u_k^\top X_i^*X_i^{*\top}u_k \le \lambda_{\max}(N^{-1}\sum_{i=1}^NX_i^*X_i^{*\top}) = O_p(1)$. As a result, we have $\sum_{i=1}^N u_k^\top X_i^*X_i^{*\top}u_k = O_p(N)$. Consequently, by Condition (C1), the third term of \eqref{vecB Taylor} is of the order $p^{1/2}N^{-3/2}O_p(\sqrt{ p\log N})O_p(N) = O_p(\sqrt{p^2\log N/N}) = O_p(\sqrt{n^2\log N/N}) = o_p(1)$.
Then we can choose a sufficiently large $C$ such that $\mL_N^{\text{Source}}(\mathbb{B}+ C \sqrt{p/N}\mathbb{U}) - \mL_N^{\text{Source}}(\mathbb{B}) <0$, which completes the theorem proof.

\csection{Proof of Theorem \ref{Theorem 2}}

The theorem conclusion can be established in three steps. In the first step, we show that there exists a local optimizer $\widehat\gamma$, which is $\sqrt n$-consistent. In the second step, we study the asymptotic normality of $\widehat\gamma$. In the third step, we analyse the asymptotic behavior of $\widehat\theta^{\text{Transfer}}$.
\scsubsection{Step 1. $\sqrt{n}$-consistency}

Recall that $\mathcal L_n(\gamma;\widehat Z) = \sum_{i=1}^n [Y_i(\widehat Z_i^\top \gamma) - \log\{1+\exp(\widehat Z_i^\top\gamma)\}] $ is the working log-likelihood. By \cite{fan2001variable}, we would like to show that, for an arbitrary but fixed $\ve >0 $, we have
\beq
\liminf_{n\to\infty}P\Big\{\sup_{\|u\| = 1} \mathcal L_n\big(\gamma + C_\ve n^{-1/2}u;\widehat Z\big) < \mathcal L_n\big(\gamma ;\widehat Z\big)\Big\} \ge 1-\varepsilon, \label{liminf probability}
\eeq
for a sufficiently large constant $C_\ve>0$. To this end, Taylor expansion is applied as
\begin{gather}
\mathcal L_n\Big(\gamma + C_\ve n^{-1/2}u;\widehat Z\Big) -\mathcal L_n\Big(\gamma ;\widehat Z\Big) \nonumber \\
= C_\ve n^{-1/2} \dot{\mL}_n(\gamma;\widehat Z)^\top u - 2^{-1}C_\ve^2 u^\top \Big\{-n^{-1}\ddot{\mL}_n(\gamma; \widehat Z)\Big\} u + o_p(1),\label{Ln Taylor}
\end{gather}
where $\dot{\mL}_n(\gamma,\widehat Z)$ and $\ddot{\mL}_n(\gamma,\widehat Z)$ are the first-order and second-order partial derivative of the working log-likelihood function $\mL_n(\gamma; \widehat Z)$ with respect to $\gamma$, respectively. To verify \eqref{liminf probability}, it suffices to show the following two key results:
\begin{gather}
(1) \sup_{\|u\| = 1} n^{-1/2} \dot{\mL}_n(\gamma;\widehat Z)^\top u = O_p(1), \label{Ln Taylor first}\\
(2) \lambda_{\min}\{-n^{-1}\ddot{\mL}_n(\gamma; \widehat Z)\} \ge \tau_{\min} > 0,\label{Ln Taylor second}
\end{gather}
for some positive constant $\tau_{\min}$ with probability tending to one. Then we verify the conclusions by the following two sub-steps.

\textsc{Step 1.1.} We start with \eqref{Ln Taylor first}. Direct computation leads to
$n^{-1/2}\dot{\mL}_n(\gamma,\widehat Z)^\top u  = n^{-1/2}\dot{\mL}_n(\gamma, Z)^\top u  + n^{-1/2}\{\dot{\mL}_n(\gamma, \widehat Z) - \dot{\mL}_n(\gamma, Z)\}^\top u$.
The first term $n^{-1/2}\dot{\mL}_n(\gamma, Z)^\top u$ can be uniformly bounded by $n^{-1/2}\| \dot{\mL}_n(\gamma,Z)\| = O_p(1)$; see \cite{fan2001variable}. We thus focus on the second term. By mean value theorem and Cauchy-Schwarz inequality, we can verify that
\begin{gather}
\bigg|n^{-1/2}\Big\{\dot{\mL}_n(\gamma, \widehat Z) - \dot{\mL}_n(\gamma, Z)\Big\}^\top u\bigg|
\le \sum_{i=1}^n \bigg|n^{-1/2}g\Big(\xi_i^\top \gamma\Big)\Big\{1-g\Big(\xi_i^\top\gamma\Big)\Big\} (\widehat Z_i - Z_i)^\top  \xi_i\gamma^\top u\bigg|\nonumber \\
+ \sum_{i=1}^n \bigg|n^{-1/2}\Big\{Y_i-g\Big(\xi_i^\top\gamma\Big)\Big\}(\widehat Z_i - Z_i)^\top u\bigg|\nonumber \\
\le \sqrt n \Big\{n^{-1}\sum_{i=1}^n\|\widehat Z_i - Z_i\|^2\Big\}^{1/2} \bigg[ n^{-1}\sum_{i=1}^n \Big\|g\Big(\xi_i^\top \gamma\Big)\Big\{1-g\Big(\xi_i^\top\gamma\Big)\Big\} \xi_i\Big\|^2\bigg]^{1/2}\|\gamma\|\nonumber \\
	 + \bigg\{\sum_{i=1}^n\Big|Y_i - g\Big(\xi_i^\top \gamma \Big)\Big|^2\bigg\}^{1/2} \Big\{n^{-1}\sum_{i=1}^n\|\widehat Z_i - Z_i\|^2\Big\}^{1/2} \label{first derivative decom},
\end{gather} where $\xi_i = \eta_i\widehat Z_i + (1-\eta_i)Z_i$ for some $\eta_i\in(0,1)$. We first evalutate the first term of \eqref{first derivative decom}. Simple computation leads to $n^{-1}\sum_{i=1}^n \|\widehat Z_i - Z_i\|^2 = n^{-1}\sum_{i=1}^n\|(\wh {\mathbb{B} }- \mathbb{B})^\top X_i\|^2 = \sum_{k=1}^K(\wh \beta_k - \beta_k)^\top  (n^{-1}\sum_{i=1}^nX_iX_i^\top)(\wh \beta_k - \beta_k) \le \lambda_{\max}(n^{-1}\sum_{i=1}^nX_iX_i^\top)  \sum_{k=1}^K\|\wh\beta_k - \beta_k\|^2$. In the meanwhile, by Theorem 6.5 in \cite{wainwright2019high} and Condition (C1), we know that there exists some constant $C_1>0$ such that $\lambda_{\max}\{n^{-1}\sum_{i=1}^{n}X_iX_i^\top - E(X_iX_i^\top )\} \le C_1(\sqrt{p/n} + p/n) \le C_1(\sqrt C + C)$ with probability tending to one. Then by Condition (C3), we have $\lambda_{\max}(n^{-1}\sum_{i=1}^{n}X_iX_i^\top ) \le C_1(\sqrt C + C) + \lambda_{\max}\{E(X_iX_i^\top)\} \le C_1(\sqrt C + C) + C_{\max}$ with probability tending to one. That is, $\lambda_{\max}(n^{-1}\sum_{i=1}^{n}X_iX_i^\top ) = O_p(1)$. By Theorem \ref{Theorem 1}, we have $\|\wh\beta_k - \beta_k\| = O_p(\sqrt{p/N})$ for $k = 1,\dots,K$. As a consequence, we have $n^{-1}\sum_{i=1}^n \|\widehat Z_i - Z_i\|^2 = O_p(p/N)$. Furthermore, note that $\xi_i = \eta_i (\widehat Z_i - Z_i) + Z_i$. By Condition (C4), we know that $E\|Z_i\|^2 \le \{E\|Z_i\|^4\}^{1/2} \le \tau_{\max}^{1/2}$. We then have $n^{-1}\sum_{i=1}^n \|g(\xi_i^\top\gamma)\{1-g(\xi_i^\top\gamma)\} \xi_i\|^2 \le n^{-1}\sum_{i=1}^n \|\xi_i\|^2 \le 2n^{-1}\sum_{i=1}^n( \|Z_i\|^2 + \|\eta_i(\widehat Z_i - Z_i)\|^2 ) = O_p(1) $. As a consequence, the first term of \eqref{first derivative decom} is of order $O_p(\sqrt{np/N})$. Then we consider the second term of \eqref{first derivative decom}. We have $\{\sum_{i=1}^n|Y_i - g(\xi_i^\top \gamma )|^2\}^{1/2} \le \sqrt n$ since $|Y_i - g(\xi_i^\top\gamma)|\le 1$. Note that $n^{-1}\sum_{i=1}^n \|\widehat Z_i - Z_i\|^2 = O_p(p/N)$. Then the second term of \eqref{first derivative decom} is of order $O_p(\sqrt{np/N})$. Combining all the preceding results, we obtain that $|n^{-1/2}\{\dot{\mL}_n(\gamma, \widehat Z) - \dot{\mL}_n(\gamma, Z)\}^\top u| = O_p(\sqrt{np/N})$. By Condition (C1) we know that $np/N\le Cn^2/N\to 0$. Then we have $n^{-1/2}\dot{\mL}_n(\gamma,\widehat Z)^\top u  = n^{-1/2}\dot{\mL}_n(\gamma, Z)^\top u  + o_p(1) = O_p(1)$. This proves \eqref{Ln Taylor first}.

\textsc{Step 1.2.} We next study \eqref{Ln Taylor second}. Recall $I (\gamma) = -E\{n^{-1}\ddot{\mL}_n(\gamma, Z)\}$ is the Fisher information matrix of the oracle likelihood. Then $\eqref{Ln Taylor second}$ follows if we can show that
\begin{gather}
\sup_{\|u\| = 1}\Big|u^\top\Big\{-n^{-1}\ddot{\mathcal{L}}_n( \gamma;\widehat Z) -  I(\gamma)\Big\}u\Big| \le \sup_{\|u\| = 1}\Big|n^{-1}u^\top \ddot{\mathcal{L}}_n( \gamma;\widehat Z) u-  n^{-1}u^\top  \ddot{\mathcal{L}}_n( \gamma;Z)u\Big| \nonumber \\
+\sup_{\|u\| = 1}\Big|u^\top\Big\{-n^{-1} \ddot{\mathcal{L}}_n(\gamma;Z)-  I(\gamma)\Big\}u\Big| = o_p(1).\label{sup decom}
\end{gather}
By Law of Large Numbers, we have $-n^{-1} \ddot{\mL}_n(\gamma; Z) \stackrel{p}{\to} I(\gamma)$. As a result, \eqref{Ln Taylor second} follows if we can show that $\sup_{\|u\| = 1}n^{-1}u^{\top}  \{\ddot{\mathcal{L}}_n( \gamma;\widehat Z) -  \ddot{\mathcal{L}}_n( \gamma;Z)\}u = o_p(1)$. By the mean value theorem, we can write
\beqrs
& & \Big|u^\top  \Big\{n^{-1}  \ddot{\mathcal{L}}_n(\gamma;\widehat Z) - n^{-1}  \ddot{\mathcal{L}}_n(\gamma;Z)\Big\}u\Big| \\
&=& n^{-1}\bigg|\sum_{i=1}^n\bigg[g\Big(\widehat Z_i^\top \gamma \Big)\Big\{1-g\Big(\widehat Z_i^\top \gamma \Big)\Big\} \Big(\widehat Z_i^\top u\Big)^2 - g\Big(Z_i^\top \gamma \Big)\Big\{1-g\Big(Z_i^\top \gamma \Big)\Big\} \Big(Z_i^\top u\Big)^2\bigg]\bigg|\\
&=&\bigg[\Big\{1-2g\Big(\tilde\xi_i^\top \gamma \Big)\Big\}g\Big(\tilde\xi_i^\top \gamma \Big)\Big\{1-g\Big(\tilde\xi_i^\top \gamma \Big)\Big\} \Big(\tilde{\xi}_i^\top u\Big)^2 \gamma + 2g\Big(\tilde{\xi}^\top \gamma\Big)\Big\{1-g\Big(\tilde{\xi}_i^\top\gamma\Big) \Big\}\Big(\tilde{\xi}_i^\top u\Big) u \bigg] \\
& \times & n^{-1}\bigg|\sum_{i=1}^n \Big(\widehat Z_i - Z_i\Big)^\top \bigg|
\le \Big\{n^{-1}\sum_{i=1}^n \|\widehat Z_i - Z_i\|^2 \Big\}^{1/2}\times\Big(I_1+I_2\Big) = O_p(\sqrt{p/N}) (I_1 + I_2),
\eeqrs
where $I_1 =  \big\{ n^{-1} \sum_{i=1}^n \big\|\big\{1-2g\big(\tilde\xi_i^\top \gamma \big)\big\}g\big(\tilde\xi_i^\top \gamma \big)\big\{1-g\big(\tilde\xi_i^\top \gamma \big)\big\} \big(\tilde{\xi}_i^\top u\big)^2 \gamma\big\|^2 \big\}^{1/2}$, $I_2 = \big\{ n^{-1} \sum_{i=1}^n \\ \big\|2g\big(\tilde{\xi}^\top \gamma\big)\big\{1-g\big(\tilde{\xi}_i^\top\gamma\big) \big\}\big(\tilde{\xi}_i^\top u\big) u\big\|^2 \big\}^{1/2}$, and $\tilde{\xi}_i = \tilde{\eta}_i \widehat Z_i + (1 - \tilde{\eta}_i) Z_i$ for some $\tilde\eta_i \in [0,1]$. Next, we should evaluate $I_1$ and $I_2$ sequentially. For $I_1$, we can verify that $
I_1^2
\le  n^{-1}\|\gamma\|^2 \sum_{i=1}^n (\tilde{\xi}_i^\top u )^2
\le n^{-1} \|\gamma\|^2 \sum_{i=1}^n \|\tilde{\xi}_i\|^2
$
The first inequality holds by the following results: (a) $|g(\tilde \xi_i^\top\tilde{\gamma})\{1-g(\tilde \xi_i^\top\tilde{\gamma})\}|\le 1$ and (b) $|1-2g(\tilde \xi_i^\top\tilde{\gamma})|\le 1$. Then by a similar procedure as in \textsc{Step 1.1}, we have $n^{-1}\sum_{i=1}^n \|\tilde{\xi}_i\|^2 = O_p(1)$. Note that by assuming $\gamma$ is bounded, we have $I_1 = O_p(1)$. For $I_2$, by similar argument above, we can show that $I_2^2 \le n^{-1} \sum_{i=1}^n \|\tilde{\xi}_i\|^2 = O_p(1)$. Thus, the first term of \eqref{sup decom} is bounded by $O_p(\sqrt {p/N}) = o_p(1)$. Combining the above results, we have $\sup_{\|u\| = 1}|-n^{-1} u^\top\ddot{\mathcal{L}}_n(\gamma;\widehat Z)u - u^\top I(\gamma)u| = o_p(1)$ and $\inf_{\|u\| = 1} \big\{-n^{-1} u^\top\ddot{\mathcal{L}}_n( \gamma;\widehat Z)u\big\} \ge \inf_{\|u\| = 1}\big\{-n^{-1} u^\top\ddot{\mathcal{L}}_n( \gamma;\widehat Z)u - u^\top I( \gamma)u\big\} \\
+ \inf_{\|u\| = 1}\big\{u^\top I(\gamma) u \big\} = o_p(1) + \lambda_{\min}\{I(\gamma) \}.$ Note that $I(\gamma)$ is a positive definite matrix whose eigenvalues are bounded away from zero. Consequently, we have $\lambda_{\min}\{I(\gamma) \}>0$. As a consequence, we have $\lambda_{\min}\{n^{-1} u^\top\ddot{\mathcal{L}}_n( \gamma;\widehat Z)u\} = \inf_{\|u\| = 1} \{n^{-1} u^\top\ddot{\mathcal{L}}_n( \gamma;\widehat Z)u\} > 2^{-1}\lambda_{\min}\{I(\gamma) \} $ with probability tending to one. Thus, for the $\ve$ mentioned above, we can take a sufficiently large $C_\ve$ such that $\sup_{\|u\| = 1} \mL_n(\gamma + C_\ve n^{-1/2}u ;\widehat Z) < \mL_n(\gamma;\widehat Z) $. Combining the preceding steps, we complete the proof of $\sqrt n$-consistency.

\scsubsection{Step 2. Asymptotic Normality}

We next study the asymptotic distribution of $\widehat \gamma$. Note that $\widehat \gamma$ is a local optimizer of $\mL_n(\gamma;\widehat Z)$. We then should have $\dot{\mL}_n(\gamma;\widehat Z) = 0$. Recall that $\widehat \gamma$ is $\sqrt n$-consistent for $\gamma$. We can then apply Taylor's expansion at $\gamma$. This leads to
\beq
0 = n^{-1/2} \dot{\mL}_n(\widehat \gamma;\widehat Z) = n^{-1/2} \dot{\mL}_n(\gamma; \widehat Z) + n^{-1}\ddot{\mL}_n(\gamma; \widehat Z) \sqrt n(\widehat \gamma - \gamma)+o_p(1). \label{normality Taylor}
\eeq
Then we verify the asymptotic normality of $\widehat \gamma $ by the following two conclusions: (a) $n^{-1/2} \dot{\mL}_n(\gamma; \widehat Z) \stackrel{d}{\to } N(0, I(\gamma))$; and (b) $-n^{-1}\ddot{\mL}_n(\gamma; \widehat Z) \stackrel{p}{\to} I(\gamma)$. We next prove the two above conclusions separately.

\textsc{Step 2.1.} We start with the asymptotic normality of $n^{-1/2} \dot{\mL}_n(\gamma;\widehat Z)$. Simple computation leads to $n^{-1/2} \dot{\mL}_n( \gamma;\widehat Z) = n^{-1/2} \dot{\mL}_n( \gamma;Z) + n^{-1/2} \{\dot{\mL}_n( \gamma;\widehat Z) -\dot{\mL}_n( \gamma;Z)\}$. By Central Limit Theorem, we have $n^{-1/2} \dot{\mL}_n( \gamma;Z) \stackrel{d}{\to} N(0,I(\gamma))$. By a similar procedure in \textsc{Step 1.1} and the property of $\ell_2$ norm, we have $n^{-1/2}\|\dot{\mL}_n( \gamma;\widehat Z) -\dot{\mL}_n( \gamma;Z)\| = \sup_{\|u\| = 1}n^{-1/2}\|\{\dot{\mL}_n( \gamma;\widehat Z) -\dot{\mL}_n( \gamma;Z)\}^\top u\| = o_p(1)$. Combining the above two results, we have $n^{-1/2} \dot{\mL}_n(\gamma; \widehat Z) \stackrel{d}{\to } N(0, I(\gamma))$.

\textsc{Step 2.2.} We show consistency of $-n^{-1}\ddot{\mL}_n(\gamma;\widehat Z)$. For an arbitrary matrix $A = (a_{ij}) \in \mR^{m\times n}$, define its spectral norm as $\|A\|_2 = \sqrt{\lambda_{\max}(A^\top A)}$. By \eqref{sup decom}, we have
\[
\Big\|-n^{-1}\ddot{\mathcal{L}}_n( \gamma;\widehat Z) -  I(\gamma)\Big\|_2 = \sup_{\|u\| = 1}\Big|u^\top\Big\{-n^{-1}\ddot{\mathcal{L}}_n( \gamma;\widehat Z) -  I(\gamma)\Big\}u\Big| = o_p(1).
\]
Consequently, we have $-n^{-1}\ddot{\mL}_n(\gamma; \widehat Z) \stackrel{p}{\to} I(\gamma)$. Then $n^{-1}\ddot{\mL}_n(\gamma; \widehat Z)$ is invertible with probability tending to one. As a result, by Slutsky's theorem, we have
$
\sqrt n(\widehat \gamma - \gamma) = \{-n^{-1}\ddot{\mL}_n(\gamma;\widehat Z) \}^{-1}\{n^{-1/2}\dot{\mL}_n(\gamma;\widehat Z) \} + o_p(1) \stackrel{d}{\to} N(0,I(\gamma)^{-1}).
$
This completes the proof of asymptotic normality of $\widehat\gamma$.

\scsubsection{Step 3. Consistency and Asymptotic Normality of $\widehat\theta^{\text{Transfer}}$}

We next study the asymptotic behavior of $\widehat\theta^{\text{Transfer}}$. The proof includes two parts. In the first part, we verify the $\sqrt n$-consistency of $\widehat\theta^{\text{Transfer}}$, while in the second part, we shown the asymptotic normality of $\widehat\theta^{\text{Transfer}}$.

\textsc{Step 3.1.} We start with the $\sqrt n$-consistency. By Cauchy-Schwarz inequality and the property of operator norm \citep{seber2008matrix}, we have $||\widehat{\theta}^{\text{Transfer}}-\theta||\leq ||\widehat{\mathbb{B}}_{\text{mle}}^{\text{Source}}-\mathbb B||_F||\widehat \gamma||+\lambda_{\max}^{1/2}(\mathbb B^\top\mathbb B)||\widehat\gamma-\gamma||$. Recall that we have shown $\widehat\gamma - \gamma = O_p(1/\sqrt n)$ in \textit{Step 1}. By Theorem \ref{Theorem 1}, the first term is of order  $O_p(\sqrt{p/N})\{\gamma+O_p(1/\sqrt{n})\}=O_p(\sqrt{p/N})$.  Then by Condition (C5), we know the second term is of order $O(1)O_p(1/\sqrt{n}) = O_p(1/\sqrt{n})$. Furthermore, given $n^2\log N/N \to 0$ and $p\le Cn$ by Condition (C1), we can obtain $||\widehat{\theta}^{\text{Transfer}}-\theta||=O_p(1/\sqrt{n})$ and thus $\widehat{\theta}^{\text{Transfer}}$ is $\sqrt n$-consistent.

\textsc{Step 3.2.} Then we evaluate the asymptotic normality of $\wh\theta^{\text{Transfer}}$. By the theorem condition $v\in\mS(\mathbb B)$ , there exists a vector $\zeta\in \mR^{K}$ such that $v = \mathbb B \zeta$. As a result, we have $\sigma^2(v) = v^\top \mathbb B I^{-1}(\gamma) \mathbb B^\top v = \zeta^\top \mathbb B^\top \mathbb B I^{-1}(\gamma) \mathbb B^\top\mathbb B \zeta \ge \lambda_{\max}^{-1}\{I(\gamma)\} \lambda_{\min}\{\mathbb B^\top \mathbb B\}$. Similarly, we have $\sigma^2(v) \le \lambda_{\min}^{-1}\{I(\gamma)\}\lambda_{\max}(\mathbb B^\top \mathbb B)$. By Condition (C5) and the positive definiteness of $I(\gamma)$ and $\mathbb B^\top \mathbb B$, we have $\sigma^2(v) = O(1)$ and $\sigma^{-2}(v) = O(1)$. Note that $\sqrt nv^\top(\wh\theta^{\text{Transfer}} - \theta)/\sigma(v) = \sqrt nv^\top(\wh{\mathbb B}_{\text{mle}}^{\text{Source}} - \mathbb B)\wh\gamma/\sigma(v) +\sqrt n v^\top \mathbb B (\wh\gamma - \gamma)/\sigma(v)$. By Condition (C1), Theorem \ref{Theorem 1}, and Cauchy-Schwarz inequality, we know that $\sqrt nv^\top(\wh{\mathbb B}_{\text{mle}}^{\text{Source}} - \mathbb B)\wh\gamma/\sigma(v)\le \sqrt n\|v\| \|\wh{\mathbb B}_{\text{mle}}^{\text{Source}} - \mathbb B\|_F \|\wh\gamma\|/\sigma(v) =\sqrt nO_p(\sqrt {p/N})O_p(1) = O_p(\sqrt{np/N}) = O_p(\sqrt {n^2/N}) = o_p(1)$. Note that $\|\mathbb B^\top v\| \le \lambda_{\max}(\mathbb B^\top \mathbb B) \le \tau_{\max}$. By the results in \textsc{Step 1.1}, \eqref{sup decom} and \eqref{normality Taylor}, we know that
\begin{gather}
0 = n^{-1/2}\sigma^{-1}(v)v^\top \mathbb B \Big\{-n^{-1}\ddot{\mL}_n(\gamma; \widehat Z)\Big\}^{-1} \dot{\mL}_n(\gamma; \widehat Z) - \sqrt n\sigma^{-1}(v)v^\top \mathbb B (\wh \gamma - \gamma)+o_p(1)\nonumber\\
= n^{-1/2}\sigma^{-1}(v)v^\top \mathbb BI(\gamma)^{-1}\dot{\mL}_n(\gamma; Z) - \sqrt n\sigma^{-1}(v)v^\top \mathbb B (\wh \gamma - \gamma) + o_p(1)\nonumber\\
= \sum_{i=1}^n Y_{ni} - \sqrt n\sigma^{-1}(v)v^\top \mathbb B (\wh \gamma - \gamma) + o_p(1)
 \label{normality of theta},
\end{gather}
where $Y_{ni}  = n^{-1/2}\sigma^{-1}(v)\{Y_i - g(Z_i^\top \gamma)\}v^\top \mathbb BI(\gamma)^{-1} Z_i$ for $1\le i\le n$. Then by the Lindeberg-Feller Central Limit Theorem \citep{vaart_1998}, the asymptotic normality of the first term of \eqref{normality of theta} holds if we can show for any $\eta_0 >0$
\begin{gather}
	(1) \ \sum_{i=1}^n E\|Y_{ni}\|^2I\{\|Y_{ni}\|\ge \eta_0\} \to 0 \quad \text{and}\quad (2) \ \sum_{i=1}^n \var(Y_{ni}) \to 1.\label{LFCLT}
\end{gather}
Then we verify these two conclusions subsequently. For the first conclusion, by Chebyshev's inequality and Condition (C5), we have
\begin{gather*}
	\sum_{i=1}^n E\|Y_{ni}\|^2I\{\|Y_{ni}\|\ge \eta_0\} \le \eta_0^{-2}\sum_{i=1}^nE\|Y_{ni}\|^4
	\le n^{-2}\sigma^{-4}(v)\sum_{i=1}^{n}E\Big\{v^\top \mathbb BI(\gamma)^{-1}Z_i\Big\}^4 \\\le n^{-2}\sigma^{-4}(v)\sum_{i=1}^{n}\lambda_{\max}^2(\mathbb B^\top \mathbb B)\lambda_{\min}^{-4}\{I(\gamma)\} E\|Z_i\|^4 =O(n^{-1}) = o(1).
\end{gather*}
As a consequence, the first conclusion of \eqref{LFCLT} holds. Then we consider the second conclusion of \eqref{LFCLT}. By direct computation, we know that $E(Y_{ni}) = n^{-1/2}\sigma^{-1}(v)v^\top \mathbb B\times I(\gamma)^{-1} E[\{Y_i - g(Z_i^\top \gamma)\}Z_i] = 0$. Then we have
\begin{gather*}
	\sum_{i=1}^{n}\var(Y_{ni}) = \sum_{i=1}^{n}n^{-1}\sigma^{-2}(v)v^\top \mathbb BI(\gamma)^{-1} E\bigg[\Big\{Y_i - g(Z_i^\top \gamma)\Big\}^2 Z_iZ_i^\top\bigg] I(\gamma)^{-1} \mathbb B^\top v = 1.
\end{gather*}
Then we have verified the second conclusion of \eqref{LFCLT}. Combining the above results, we have $\sqrt n\sigma^{-1}(v)v^\top(\wh\theta^{\text{Transfer}} - \theta) = \sqrt n\sigma^{-1}(v)v^\top \mathbb B (\wh \gamma - \gamma) +o_p(1)
	= n^{-1/2}\sigma^{-1}(v)v^\top \mathbb BI(\gamma)^{-1}\\\dot{\mL}_n(\gamma; Z) + o_p(1) \stackrel{d}{\to}N(0,1)$.
Then we complete the proof of asymptotic normality of $\wh{\theta}^{\text{Transfer}}$. The theorem proof is complete.

\csection{Proof of Theorem \ref{Theorem 3}}

Define $\widetilde \mW =\diag[g(\widehat Z_1^\top\gamma)\{1-g(\widehat Z_1^\top\gamma)\},\dots,g(\widehat Z_n^\top\gamma)\{1-g(\widehat Z_n^\top\gamma)\} ]\in \mR^{n\times n}$ and $\mW = \diag[g(Z_1^\top\gamma)\{1-g(Z_1^\top\gamma)\},\dots,g(Z_n^\top\gamma)\{1- g(Z_n^\top\gamma)\} ]\in \mR^{n\times n}$. Let $\mZ = (Z_1,\dots,Z_n)^\top \in \mR^{n\times K}$ be the design matrix of the dimension reduced features and $\widehat\mZ = (\widehat Z_1,\dots,\widehat Z_n)^\top\in\mR^{n\times K}$ be the estimated matrix of $\mZ$. Let $\tilde\ve_i = Y_i - g(\widehat Z_i^\top\gamma)$ and $\tilde{\ve} = (\tilde{\ve}_1,\dots,\tilde{\ve}_n)^\top\in\mR^n$. Recall that when the null hypothesis holds, the pseudo residual is $\ve_i = Y_i - g(Z_i^\top\gamma)$. Let $\ve = (\ve_1,\dots,\ve_n)^\top \in\mR^n$ be the vector of pseudo residuals. Then by the proof in Theorem \ref{Theorem 2}, we have $\wh \gamma - \gamma = (\widehat \mZ^\top\widetilde \mW\widehat \mZ)^{-1} \widehat \mZ^\top \tilde\ve +o_p(n^{-1/2})$. Recall that $\wh\ve = (\wh\ve_1,\dots,\wh\ve_n)^\top $ with $\wh \ve_i = Y_i - g(\widehat Z_i^\top\wh\gamma)$. By Taylor expansion, we have
\[
\widehat\ve = \tilde{\ve} - \widetilde\mW\widehat\mZ(\widehat\gamma - \gamma) + o_p(n^{-1/2}) = \Big\{\mI_n - \widetilde\mW\widehat\mZ\Big(\widehat \mZ^\top\widetilde \mW\widehat \mZ\Big)^{-1} \widehat \mZ^\top\Big\}\tilde{\ve} + o_p(n^{-1/2}),
\]
where $\mI_n$ is the identity matrix. Then the test statistic can be written as
\begin{gather*}
T_1 = n^{-2}\tilde{\ve}^\top\Big\{\mI_n - \widehat\mZ\Big(\widehat \mZ^\top\widetilde \mW\widehat \mZ\Big)^{-1} \widehat \mZ^\top\widetilde\mW\Big\}\mX \mX^\top\Big\{\mI_n - \widetilde\mW\widehat\mZ\Big(\widehat \mZ^\top\widetilde \mW\widehat \mZ\Big)^{-1} \widehat \mZ^\top\Big\}\tilde\ve +o_p(n^{-1})\\
= n^{-2}\Big\{\ve^\top \mX\mX^\top\ve + (\tilde\ve^\top\mX\mX^\top\tilde\ve - \ve^\top\mX\mX^\top\ve) + \tilde\ve^\top \widehat\mZ\Big(\widehat \mZ^\top\widetilde \mW\widehat \mZ\Big)^{-1} \widehat \mZ^\top\widetilde\mW \mX \mX^\top\widetilde\mW\widehat\mZ\Big(\widehat \mZ^\top\widetilde \mW\widehat \mZ\Big)^{-1} \widehat \mZ^\top\tilde\ve\\
-2\tilde\ve^\top \mX \mX^\top\widetilde\mW\widehat\mZ\Big(\widehat \mZ^\top\widetilde \mW\widehat \mZ\Big)^{-1} \widehat \mZ^\top\tilde\ve \Big\}+o_p(n^{-1}) = T_1^* + R_1 + R_2 -2 R_3 +o_p(n^{-1}),
\end{gather*}
where $T_1^* = n^{-2}\ve^\top \mX\mX^\top\ve, R_1 = n^{-2} (\tilde\ve^\top\mX\mX^\top\tilde\ve - \ve^\top\mX\mX^\top\ve), R_2 = n^{-2}\tilde\ve^\top \widehat\mZ(\widehat \mZ^\top\widetilde \mW\widehat \mZ)^{-1} \widehat \mZ^\top\widetilde\mW \times\mX \mX^\top\widetilde\mW\widehat\mZ(\widehat \mZ^\top\widetilde \mW\widehat \mZ)^{-1} \widehat \mZ^\top\tilde\ve$, and $R_3 = n^{-2}\tilde\ve^\top\mX \mX^\top\widetilde\mW\widehat\mZ(\widehat \mZ^\top\widetilde \mW\widehat \mZ)^{-1} \widehat \mZ^\top\tilde\ve$. Recall that $\Sigma_\gamma = E(\ve_i^2X_iX_i^\top) = E[g(Z_i^\top\gamma)\{1-g(Z_i^\top\gamma)\}X_iX_i^\top ] = (\sigma_{j_1j_2}^{(\gamma)})_{p\times p} $. By Condition (C3), we know that $pC_{\min}^2\le \tr(\Sigma_\gamma^2) \le pC_{\max}^2$. Then, the theorem conclusion follows if we can show the following five results
\begin{gather}
	(1) E(T_1^*) = n^{-1}\tr(\Sigma_\gamma),\label{thm3con1}\\
	(2) \var(T_1^*) = n^{-3}\var(\ve_i^2 X_i^\top X_i) +  2n^{-2}\tr(\Sigma_\gamma^2)\{1+o(1)\}, \label{thm3con2}\\
	(3) R_1 = o_p\{n^{-1}\sqrt p\},\label{thm3con3} \\
	(4) R_2 = o_p\{n^{-1}\sqrt p\},\label{thm3con4} \\
	 (5) R_3 = o_p\{n^{-1}\sqrt p\} \label{thm3con5}.
\end{gather}
Note that if the above results hold, we should have $R_i / \var^{1/2}(T_1^*) \stackrel{p}{\to}0$ for $i=1,2,3$. The remainder terms should be negligible. The five conclusions are to be verified subsequently.

\textsc{Step 1.} We start with $E(T_1^*)$. Under the null hypothesis, we know that conditional on $\mZ$, $Y_i$ is independent with $X_i$. Thus, we can compute the conditional expectation as $E(\ve^\top\mX\mX^\top\ve |\mZ) = E\{\tr(\ve\ve^\top\mX\mX^\top)|\mZ\}
= \tr\{E(\ve\ve^\top|\mZ)E(\mX\mX^\top|\mZ) \}
  = \tr\{\mW E(\mX\mX^\top|\mZ) \}
  = \tr\{E(\mX^\top\mW\mX|\mZ)\} = \tr\{E[\sum_{i=1}^ng(Z_i^\top\gamma)\{1-g(Z_i^\top\gamma)\}X_iX_i^\top |\mZ]\} = n\tr\{E[g(Z_i^\top\gamma)\{1-g(Z_i^\top\gamma)\}X_iX_i^\top|\mZ]\}$. Thus, we have $E(T_1^*) = n^{-2} E\{E(\ve^\top\mX\mX^\top\ve |\mZ) \}
  = n^{-1}\tr(\Sigma_\gamma)$.

\textsc{Step 2.} We consider $\var(T_1^*) = n^{-4} E(\ve^\top\mX\mX^\top\ve)^2 - n^{-4}\{E(\ve^\top\mX\mX^\top\ve)\}^2$. We first consider $E(\ve^\top\mX\mX^\top\ve)^2$. Then we have
\beqrs
	&&E\Big(\ve^\top\mX\mX^\top\ve\Big)^2 = E\bigg\{\sum_{j=1}^p\Big(\sum_{i=1}^n X_{ij}\ve_i\Big)^2 \bigg\}^2 = E\bigg(\sum_{j=1}^p\sum_{i_1,i_2} \ve_{i_1}\ve_{i_2}X_{i_1j}X_{i_2j} \bigg)^2 \\
	&=& \sum_{j_1,j_2}\sum_{i_1i_2i_3i_4} E\Big(\ve_{i_1}\ve_{i_2}\ve_{i_3}\ve_{i_4}X_{i_1j_1}X_{i_2j_2}X_{i_3j_1}X_{i_4j_2}\Big)\\
	&=&\sum_{j_1,j_2} \sum_{i=1}^n E\Big(\ve_i^4X_{ij_1}^2X_{ij_2}^2\Big) + \sum_{j_1,j_2}\sum_{i_1\neq i_2}E\Big(\ve_{i_1}^2\ve_{i_2}^2X_{i_1j_1}^2X_{i_2j_2}^2\Big)\\
	&+&2\sum_{j_1,j_2}\sum_{i_1\neq i_2}E\Big(\ve_{i_1}^2\ve_{i_2}^2 X_{i_1j_1}X_{i_1j_2}X_{i_2j_1}X_{i_2j_2} \Big)\\
	&=& nE\bigg\{\ve_i^4\Big(X_i^\top X_i\Big)^2\bigg\} + n(n-1)\sum_{j_1,j_2}E\Big(\ve_{i}^2X_{ij_1}^2\Big)E\Big(\ve_{i_2}^2X_{ij_2}^2\Big) \\
	&+& 2n(n-1)\sum_{j_1,j_2}E^2\Big(\ve_i^2X_{ij_1}X_{ij_2}\Big)\\
	&=& nE\bigg\{\ve_i^4\Big(X_i^\top X_i\Big)^2\bigg\} + n(n-1)\bigg\{\sum_{j}\sigma_{jj}^{(\gamma)}\bigg\}^2
	+ 2n(n-1)\sum_{j_1,j_2}\sigma_{j_1j_2}^{2(\gamma)}\\
	&=&nE\bigg\{\ve_i^4\Big(X_i^\top X_i\Big)^2\bigg\} + n(n-1)\tr^2(\Sigma_\gamma)
	+ 2n(n-1)\tr(\Sigma_\gamma^2).
\eeqrs
Then we have $\var(T_1^*) = n^{-4} E(\ve^\top\mX\mX^\top\ve)^2 - n^{-4}\{E(\ve^\top\mX\mX^\top\ve)\}^2 = n^{-3}E\{\ve_i^4(X_i^\top X_i)^2\} - n^{-3}\tr^2(\Sigma_\gamma) + 2n^{-2}(1-n^{-1})\tr(\Sigma_\gamma^2) = n^{-3}\var(\ve_i^2X_i^\top X_i) + 2n^{-2}(1-n^{-1})\tr(\Sigma_\gamma^2)$. 

\textsc{Step 3.} Next we study the order of $R_1$. By the mean value theorem, we have
\beqrs
	|R_1| &=& n^{-2} \Big|\tilde\ve^\top\mX\mX^\top\tilde\ve - \ve^\top\mX\mX^\top\ve\Big| = n^{-2} \sum_{i=1}^n \sum_{j=1}^n\Big| \Big(\tilde \ve_{i}\tilde\ve_{j} - \ve_{i}\ve_{j}\Big) X_{i}^\top X_{j}\Big|\\
	&\le& n^{-2}\sum_{i=1}^n \sum_{j=1}^n \Big|\Big(\tilde \ve_{i}\tilde\ve_{j} - \tilde \ve_{i}\ve_{j}\Big) X_{i}^\top X_{j}\Big| + n^{-2}\sum_{i=1}^n \sum_{j=1}^n \Big|\Big(\tilde \ve_{i}\ve_{j} - \ve_{i}\ve_{j}\Big) X_{i}^\top X_{j}\Big|\\
	&\le& 2n^{-2} \sum_{i=1}^n\sum_{j=1}^n \Big|(\tilde \ve_i - \ve_i )X_i^\top X_j\Big| \\
	&=& 2n^{-2}\sum_{i=1}^n\sum_{j=1}^n g\Big(\bar\xi_i^\top\gamma\Big)\Big\{1-g\Big(\bar\xi_i^\top\gamma\Big)\Big\}\Big|(\wh Z_i-Z_i)^\top\gamma X_i^\top X_j\Big|\le R_{11}+R_{12},
\eeqrs
where $\bar\xi_i = \bar{\eta}_i \widehat{Z}_i + (1-\bar{\eta}_i) Z_i$ for some $\bar{\eta}_i \in [0,1]$, $R_{11} = 2n^{-2}\sum_{i=1}^n|(\wh Z_i-Z_i)^\top\gamma X_i^\top X_i|$ and $R_{12} =2n^{-2}\sum_{i\neq j}|(\wh Z_i-Z_i)^\top\gamma X_i^\top X_j|$. Then we evaluate the above two terms subsequently. For $R_{11}$, by Cauchy-Schwarz inequality, we have $R_{11} \le 2n^{-1}\|\gamma\| \big(n^{-1} \sum_{i=1}^n \|\wh Z_i - Z_i\|^2\big)^{1/2}\big\{n^{-1} \sum_{i=1}^{n} (X_i^\top X_i)^2  \big\}^{1/2}$. By the proof of Theorem \ref{Theorem 2}, we have $n^{-1}\sum_{i=1}^{n}\|\wh Z_i - Z_i\|^2 = O_p(p/N)$. By the sub-Gaussianity of $X_i$, we have $E\{n^{-1}\sum_{i=1}^n (X_i^\top X_i)^2 \} = E(X_i^\top X_i)^2\le \kappa_{\max} p^2$ for some $0<\kappa_{max}<\infty$. Thus, we have $n^{-1} \sum_{i=1}^{n} (X_i^\top X_i)^2 = O_p(p^2)$. Note that by Condition (C1), we have $p^2/N \le n^2/ N\to 0$. As a result, we have $R_{11} = 2n^{-1}O_p(p^{1/2}/N^{1/2})O_p(p) =O_p(n^{-1}p^{3/2}N^{-1/2}) = o_p(n^{-1}p^{1/2})$.

We then focus on $R_{12}$. By Cauchy-Schwarz inequality, we have $R_{12} \le 2 \|\gamma\| \big(n^{-1} \sum_{i=1}^n \\ \|\wh Z_i - Z_i\|^2\big)^{1/2}\big\{n^{-2} \sum_{i\neq j} (X_i^\top X_j)^2  \big\}^{1/2}$.
By Condition (C4) and Cauchy-Schwarz inequality, we have $E\{n^{-2} \sum_{i\neq j} (X_i^\top X_j)^2\} = (1-n^{-1}) E[\{n(n-1)\}^{-1} \sum_{i\neq j} (X_i^\top X_j)^2] =(1-n^{-1}) E(X_i^\top X_j)^2\le \{E(X_i^\top X_j)^4\}^{1/2}\le \tau_{\max}^{1/2}p$. Note that $n^{-1}\sum_{i=1}^{n}\|\wh Z_i - Z_i\|^2 = O_p(p/N)$. As a result, we have $R_{12} = O_p(p^{1/2}N^{-1/2})O_p(p^{1/2}) = O_p(pN^{-1/2}) = o_p(n^{-1}p^{1/2})$ where the last equality from the assumption $n^2p/N\to0$. Combining the above results, the proof of \eqref{thm3con3} is completed.

\textsc{Step 4.} Next we evaluate $R_2$. Let $\wmH = \wh \mZ(\wh \mZ^\top \widetilde{\mW} \wh \mZ )^{-1} \wh \mZ^\top = (\widehat h_{ij})_{K\times K}$ and $\mathbb H = \mZ(\mZ^\top \mW \mZ)^{-1}\mZ^\top = (h_{ij})_{K\times K}$. Then $R_{2}= n^{-2}\tilde{\ve}^\top\wmH\widetilde{\mW}\mX\mX^\top\widetilde{\mW}\wmH\tilde\ve\le n^{-2}\lambda_{\max}(\mX\mX^\top)\tilde\ve^\top \wmH^2 \tilde\ve$. Next we should study $\lambda_{\max}(\mX\mX^\top)$ and $\tilde{\ve}^\top\wmH^2\tilde{\ve}$ separately.

\textsc{Step 4.1.} We first consider $\lambda_{\max}(\mX\mX^\top)$. By Chebychev inequality, we know that for any $t>0$, we have $P(\lambda_{\max}(\mX\mX^\top ) > t)\le t^{-4}E[\lambda^4_{\max}\{(\mX\mX^\top)^4\}]\le t^{-4} E[\tr\{(\mX\mX^\top)^4\}]$. Then we consider $E[\tr\{(\mX\mX^\top)^4\}]$. We decompose the index set into the following six subsets: (i) $\mathcal A_1 = \{(i_1,i_2,i_3,i_4):i_1\neq i_2\neq i_3\neq i_4\}$, (ii) $\mathcal A_2 = \{(i_1,i_2,i_3,i_4): i_{j_1} \neq i_{j_2} \neq i_{j_3} = i_{j_4} \text{ for some }j_1,j_2,j_3,j_4\in \{1,2,3,4\} \}$, (iii) $\mathcal A_3 = \{(i_1,i_2,i_3,i_4): i_1 = i_2 \neq i_3 = i_4 \text{ or }i_1 = i_4 \neq i_2 = i_3\}$, (iv) $\mathcal A_4 = \{(i_1,i_2,i_3,i_4): i_1 = i_3 \neq i_2 = i_4 \}$, (v) $\mathcal A_5 = \{(i_1,i_2,i_3,i_4): i_{j_1} = i_{j_2} = i_{j_3} \neq i_{j_4} \text{ for some }j_1,j_2,j_3,j_4\in \{1,2,3,4\} \}$ and (vi) $\mathcal A_6 = \{(i_1,i_2,i_3,i_4): i_1=i_2=i_3=i_4 \}$. Note that $E[\tr\{(\mX\mX^\top)^4\}] = \sum_{k=1}^6\sum_{\mathcal A_k}E(X_{i_1}^\top X_{i_2}X_{i_2}^\top X_{i_3}X_{i_3}^\top X_{i_4}X_{i_4}^\top X_{i_1})$.
We then evaluate the sum of the expectations in different $\mathcal A_k$'s sequentially.
Note that for $(i_1,i_2,i_3,i_4) \in \mathcal A_1$, we should have $X_{i_1},X_{i_2},X_{i_3}$ and $X_{i_4}$ mutually independent. Therefore, we have
\begin{gather*}
	\sum_{\mathcal A_1}E\Big(X_{i_1}^\top X_{i_2}X_{i_2}^\top X_{i_3}X_{i_3}^\top X_{i_4}X_{i_4}^\top X_{i_1}\Big) = \sum_{i_1\neq i_2\neq i_3\neq i_4}E\Big\{\tr\Big(X_{i_1}X_{i_1}^\top X_{i_2}X_{i_2}^\top X_{i_3}X_{i_3}^\top X_{i_4}X_{i_4}^\top \Big)\Big\}\\
	=n(n-1)(n-2)(n-3) \tr\Big\{E^4\Big(X_iX_i^\top\Big)\Big\} = O(n^4 p).
\end{gather*}

For $\mathcal A_2$, we have $|\mathcal A_2| \le |\mathcal A^c| = n^4 - |\mathcal A| = n^{4} - n(n-1)(n-2)(n-3) \le 6n^3$ when $n\ge 2$. Note that $\tr(AB) \le \tr(A)\tr(B)$ for any positive definite matrices $A$ and $B$ \citep{seber2008matrix}. Note that by sub-Gaussianity of $X_i$, we have $E(X_i^\top X_i)^4 \le \kappa_{\max}' p^4$ for some $0<\kappa_{max}'<\infty$. Then by Cauchy-Schwarz inequality, Holder's inequality, and Condition (C3), we have
\begin{equation}
\begin{split}
&\sum_{\mathcal A_2}E\Big(X_{i_1}^\top X_{i_2}X_{i_2}^\top X_{i_3}X_{i_3}^\top X_{i_4}X_{i_4}^\top X_{i_1}\Big)
	\le 6n^3 E\bigg[\tr\Big\{X_{i_1}X_{i_1}^\top X_{i_2}X_{i_2}^\top \big(X_{i_3}X_{i_3}^\top\big)^2\Big\}\bigg]\\
	=&6n^3\tr\Big\{E^2\big(X_{i}X_{i}^\top\big)E\big(X_{i}X_{i}^\top\big)^2\Big\}
	\le 6n^3 \tr^{1/2}\Big\{E^4\big(X_{i}X_{i}^\top\big)\Big\}\tr^{1/2}\Big\{E^2\big(X_{i}X_{i}^\top\big)^2\Big\}\\
	\le &6C_{\max}^2n^3p^{1/2}E^{1/2}\Big(X_{i_2}^\top X_{i_1} X_{i_1}^\top X_{i_1} X_{i_1}^\top X_{i_2} X_{i_2}^\top X_{i_2}\Big)\\
	\le& 6C_{\max}^2n^3p^{1/2} \Big\{E\big(X_{i_1}^\top X_{i_2}\big)^4\Big\}^{1/4} \Big\{E\big(X_i^\top X_i\big)^4 \Big\}^{1/4} = O(n^3p^2). \nonumber
\end{split}
\end{equation}

For $\mathcal A_3$, note that $|\mathcal A_3| \le 2n^2$. Then by the preceding result, we have
\begin{gather*}
\sum_{\mathcal A_3}E\Big(X_{i_1}^\top X_{i_2}X_{i_2}^\top X_{i_3}X_{i_3}^\top X_{i_4}X_{i_4}^\top X_{i_1}\Big) \le 2n^2 \tr\Big[E\Big\{\big(X_{i_1}X_{i_1}^\top \big)^2\big(X_{i_2}X_{i_2}^\top\big)^2\Big\}\Big] \\
= 2n^2\tr\Big\{E^2\big(X_iX_i^\top\big)^2\Big\}  \le 2C_{\max}^2n^2p^3 = O(n^2p^3).
\end{gather*}

For $\mathcal A_4$, note that $|\mathcal A_4|\le n^2$. We then have $
	\sum_{\mathcal A_4}E(X_{i_1}^\top X_{i_2}X_{i_2}^\top X_{i_3}X_{i_3}^\top X_{i_4}X_{i_4}^\top X_{i_1})  \le n^2 E(X_{i_1}^\top X_{i_2})^4 \le \tau_{max}n^2p^2 = O(n^2p^2),
$ where the last inequality holds by Condition (C3).
For $\mathcal A_5$, by Cauchy-Schwarz inequality and Condition (C4), we have
\begin{gather*}
	\sum_{\mathcal A_5} E\Big(X_{i_1}^\top X_{i_2}X_{i_2}^\top X_{i_3}X_{i_3}^\top X_{i_4}X_{i_4}^\top X_{i_1}\Big) \le n^2 E\Big\{\big(X_{i_1}^\top X_{i_1}\big)^2 \big(X_{i_1}^\top X_{i_2}\big)^2\Big\} \\
	\le n^2\Big\{E\big(X_{i}^\top X_{i}\big)^4\Big\}^{1/2} \Big\{E\big(X_{i_{1}}^\top X_{i_2}\big)^4\Big\}^{1/2}\le \kappa_{\max}'^{1/2}\tau_{\max}^{1/2}n^2p^3 = O(n^2p^3).
\end{gather*}

For $\mathcal A_6$, we know that $\sum_{\mathcal A_6} E(X_{i_1}^\top X_{i_2}X_{i_2}^\top X_{i_3}X_{i_3}^\top X_{i_4}X_{i_4}^\top X_{i_1}) = nE(X_i^\top X_i)^4 = O(np^4)$. Combining the preceding results and Condition (C1), we have
$
E[\tr\{(\mX\mX^\top)^4\}] = O(n^4p) + O(n^3p^2) + O(n^2p^3) + O(np^4)= O(n^4p).
$
As a result, we have $\lambda_{\max}(\mX\mX^\top) = O_p(np^{1/4})$.


\textsc{Step 4.2.} We study $\tilde{\ve}^\top \wmH^2 \tilde{\ve} = R_{21} + R_{22} + R_{23}$, where $R_{21} = \tilde{\ve}^\top(\wmH^2 - \mathbb H^2)\tilde{\ve}$, $R_{22} = \tilde{\ve}^\top\mathbb H^2 \tilde{\ve} - \ve^\top \mathbb H^2 \ve$ and $R_{23} = \ve^\top \mathbb H^2 \ve$. Then we evaluate the above three terms subsequently.

\textsc{Step 4.2.1.} For $R_{21}$, we first consider $\|\wmH - \mathbb H\|_F$. Note that $\|AB\|_F \le \|A\|_F\|B\|_F$ for arbitrary matrices $A$ and $B$ \citep{seber2008matrix}. Thus, we have $\|\wmH - \mathbb H\|_F \le \mathbb H_1 + \mathbb H_2$, where $
\mathbb H_1 = \tr(\widehat\mZ^\top\widehat\mZ) \|(\widehat{\mZ}^\top \widetilde{\mW} \widehat{\mZ})^{-1} - (\mZ^\top \mW \mZ)^{-1}\|_F$ and $ \mathbb H_2 = \|\widehat{\mZ}(\mZ^\top \mW\mZ)^{-1} \widehat{\mZ} -\mZ(\mZ^\top \mW\mZ)^{-1} \mZ^\top \|_F$.
Then we consider $\mathbb H_1$. By the proof in Theorem \ref{Theorem 2}, we have $\|\widehat\mZ^\top \widetilde{\mW} \widehat{\mZ} - \mZ^\top \mW \mZ \|_F^2 =  O_p(np/N)$. Then we have $\|(\widehat{\mZ}^\top \widetilde{\mW} \widehat{\mZ})^{-1} - (\mZ^\top \mW \mZ)^{-1}\|_F \le \|(\widehat{\mZ}^\top\widetilde{\mW}\widehat{\mZ})^{-1}\|_F\|\widehat\mZ^\top \widetilde{\mW} \widehat{\mZ} - \mZ^\top \mW \mZ \|_F \|(\mZ^\top \mW\mZ)^{-1}\|_F = O_p(\sqrt{p/n^3N})$ since $n^{-1}\widehat{\mZ}^\top\widetilde{\mW}\widehat{\mZ} = O_p(1)$ and $n^{-1} \mZ^\top \mW\mZ = O_p(1)$ by the proof in Theorem \ref{Theorem 2}. Note that $\tr(\widehat\mZ^\top\widehat{\mZ}) = \sum_{i=1}^n \widehat{Z}_i^\top\widehat{Z}_i = O(n)$. Thus we have $\mathbb H_1 = O_p(\sqrt{p/nN})$. Next we study $\mathbb H_2$ and have
\begin{gather*}
	\mathbb H_2 = \|\widehat{\mZ}(\mZ^\top \mW\mZ)^{-1} \widehat{\mZ} -\mZ(\mZ^\top \mW\mZ)^{-1} \mZ^\top \|_F \\\le \|\widehat{\mZ}(\mZ^\top \mW\mZ)^{-1} \widehat{\mZ} -\widehat{\mZ}(\mZ^\top \mW\mZ)^{-1}\mZ^\top \|_F + \|\widehat{\mZ}(\mZ^\top \mW\mZ)^{-1}\mZ^\top - \mZ(\mZ^\top \mW\mZ)^{-1} \mZ^\top\|_F\\
	\le  \|\widehat{\mZ}\|_F\|(\mZ^\top\mW\mZ)^{-1}\|_F \|\widehat{\mZ}-\mZ\|_F+ \|\widehat{\mZ}-\mZ\|_F\|(\mZ^\top\mW\mZ)^{-1}\|_F\|\mZ\|_F = O_p(\sqrt{p/N}),
\end{gather*}
where the last equality holds by the following facts: (1) $\|\widehat\mZ\|_F = O_p(\sqrt{n})$, (2) $\|\mZ\|_F = O_p(\sqrt{n})$, (3) $\|(\mZ^\top\mW\mZ)^{-1}\|_F = O_p(1/n)$ and (4) $\|\widehat{\mZ} - \mZ\|_F = (\sum_{i=1}^{n}\|\widehat Z_i - Z_i\|^2)^{1/2} = O_p(\sqrt{np/N})$. Combining the above results, we have $\|\wmH - \mathbb H\|_F = O_p(\sqrt{p/N})$. 
By Law of Large Numbers, we have $n^{-1}\mZ^\top \mathbb W \mZ  \stackrel{p}{\to} I(\gamma)$ and $n^{-1} \mZ^\top \mZ = n^{-1}\sum_{i=1}^nZ_iZ_i^\top \stackrel{p}{\to} E(Z_iZ_i^\top)$. Thus, we have $n^{-1}\mZ^\top \mathbb W \mZ = O_p(1)$ and $n^{-1} \mZ^\top \mZ = O_p(1)$. Note that $I(\gamma)$ is positive definite matrix with $\lambda_{\min}\{I(\gamma)\}>0$. As a result, $\lambda_{\min}(n^{-1}\mZ^\top \mathbb W \mZ) \ge 2^{-1}\lambda_{\min}\{I(\gamma)\}>0$ with probability tending to one. Then we have
\begin{gather*}
\|\mathbb H\|_F^2 = \tr(\mathbb H^2) = \tr\Big\{\mZ\Big(\mZ^\top \mathbb W \mZ\Big)^{-1}\mZ^\top\mZ\Big(\mZ^\top \mathbb W \mZ\Big)^{-1}\mZ^\top \Big\} \\
= \tr\Big\{\Big(n^{-1}\mZ^\top\mZ\Big)\Big(n^{-1}\mZ^\top \mathbb W \mZ\Big)^{-1}\Big(n^{-1}\mZ^\top\mZ\Big)\Big(n^{-1}\mZ^\top \mathbb W \mZ\Big)^{-1} \Big\} = O_p(1)
\end{gather*}
Similarly, we have $\|\wmH\|_F^2 = O_p(1)$. Then $\|\wmH^2 - \mathbb H^2\|_F \le \|\wmH - \mathbb H\|_F(\|\wmH\|_F + \|\mathbb H\|_F) = O_p(\sqrt{p/N})$. Thus we have $\tilde{\ve}^\top(\wmH^2 - \mathbb H^2)\tilde{\ve} \le \|\wmH^2 - \mathbb H^2\|_F \tilde{\ve}^\top\tilde{\ve} = O_p(\sqrt{n^2p/N}) = o_p(1)$.

\textsc{Step 4.2.2.} We study $R_{22}$. By Cauchy-Schwarz inequality, we have
\beqrs
	|R_{22}| &=& \Big|\sum_{i,j}^{n} h_{ij} \Big(\tilde{\ve}_i\tilde{\ve}_j - \ve_i \ve_j\Big) \Big|= \Big|\sum_{i,j}^{n} h_{ij} \tilde{\ve}_i\Big(\tilde{\ve}_j -\ve_j\Big) + \sum_{i,j}^{n} h_{ij}\ve_j \Big(\tilde{\ve}_i - \ve_i\Big)\Big| \\
	&\le& \Big|\sum_{i,j}^{n}h_{ij}\tilde{\ve}_i g\Big(\bar{\xi}_j^\top \gamma\Big) \Big\{1-g\Big(\bar{\xi}_j^\top\gamma\Big)\Big\}\Big(\widehat{Z}_j - Z_j\Big)^\top \gamma\Big| \\
	&+& \Big|\sum_{i,j}^{n}h_{ij}\ve_j g\Big(\bar{\xi}_i^\top \gamma\Big) \Big\{1-g\Big(\bar{\xi}_i^\top\gamma\Big)\Big\}\Big(\widehat{Z}_i - Z_i\Big)^\top \gamma\Big|\\
	&\le& 2 \Big(\sum_{i,j}^{n}h_{ij}^2\Big)^{1/2}\Big(n\sum_{i=1}^{n}\|\widehat{Z}_i - Z_i\|^2\Big)^{1/2}\|\gamma\|.
\eeqrs
By the preceding proof, we know that $\sum_{i,j}^{n}h_{ij}^2 = \tr(\mathbb H^2) = O_p(1)$. As a result, by condition $n^2p/N \to 0$, we have $|R_{22}| = O_p(1) O_p(\sqrt{n^2p/N}) = O_p(\sqrt{n^2p/N}) = o_p(1)$.

\textsc{Step 4.2.3.} We consider $R_{23}$. By direct computation, we have $R_{23} = \ve^\top \mathbb H^2\ve = \tr(\ve\ve^\top \mathbb H^2) \le \tr(\mathbb H^2) = O_p(1)$. Combining the results in {\sc Step 4.2.1} to {\sc Step 4.2.3}, we know that $\tilde{\ve}^\top \wmH^2\tilde{\ve} = O_p(1)$. Combining the preceding results, we have $
np^{-1/2}R_2 \le n^{-1}p^{-1/2}\lambda_{\max}(\mX\mX^\top ) \tilde{\ve}^\top \wmH^2\tilde{\ve} = n^{-1}p^{-1/2} O_p(n p^{1/4})O_p(1)= O_p(p^{-1/4}) = o_p(1)$. Then $R_2 = o_p(n^{-1}\sqrt p)$. We complete the proof of \eqref{thm3con4}.

\textsc{Step 5.} Now we evaluate $R_3$. Note that we have shown $E[\tr\{(\mX\mX^\top)^4\}] = O(n^4p)$ and $\|\wmH\|_F = O_p(1)$. By Cauchy-Schwarz inequality, we have
\begin{gather*}|R_3| = n^{-2}\Big|\tilde{\ve}^\top \mX\mX^\top \widetilde{\mW}\wmH\tilde\ve\Big|  = n^{-2} \Big|\tr\Big(\tilde\ve\tilde\ve^\top \mX\mX^\top \wt \mW \wmH\Big)\Big|\le n^{-2} \Big|\tr\Big(\mX\mX^\top\wmH\Big)\Big| \\\le n^{-2}\tr^{1/2}\Big\{\big(\mX\mX^\top\big)^2\Big\}\tr^{1/2}\big(\wmH^2\big) \le
n^{-2}\tr^{1/4}\Big\{\big(\mX\mX^\top\big)^4\Big\}\tr^{1/2}\big(\wmH^2\big) \\= n^{-2}O_p(np^{1/4})O_p(1) = O_p(n^{-1}p^{1/4}) = o_p(n^{-1}p^{1/2}).
\end{gather*} Then we prove the conclusion \eqref{thm3con5}. As a consequence, we complete the whole proof of the theorem.


\csection{Proof of Theorem \ref{Theorem 4}}

Define $\mathbb D_X = \diag\{X_1^\top X_1,\dots,X_n^\top X_n\}\in \mR^{n\times n}$. Then we have $\tr(\widehat\Sigma_\gamma) = n^{-1} \widehat\ve^\top \mathbb D_X\widehat\ve$. By a similar procedure of the proof in Theorem \ref{Theorem 3}, we decompose the test statistic $T_2 = T_1 - n^{-1}\tr(\widehat\Sigma_\gamma)$ as
\begin{gather*}
	T_2 = n^{-2}\Big[\ve^\top (\mX\mX^\top - \mathbb D_X)\ve + \{\tilde\ve^\top(\mX\mX^\top - \mathbb D_X)\tilde\ve - \ve^\top(\mX\mX^\top - \mathbb D_X)\ve\} \\
	+ \tilde\ve^\top \widehat\mZ\Big(\widehat \mZ^\top\widetilde \mW\widehat \mZ\Big)^{-1} \widehat \mZ^\top\widetilde\mW (\mX \mX^\top - \mathbb D_X)\widetilde\mW\widehat\mZ\Big(\widehat \mZ^\top\widetilde \mW\widehat \mZ\Big)^{-1} \widehat \mZ^\top\tilde\ve\\
	-2\tilde\ve^\top (\mX \mX^\top- \mathbb D_X)\widetilde\mW\widehat\mZ\Big(\widehat \mZ^\top\widetilde \mW\widehat \mZ\Big)^{-1} \widehat \mZ^\top\tilde\ve \Big]+o_p(1) = T_2^* + \tilde R_1 + \tilde R_2 -2 \tilde R_3 +o_p(n^{-1}),
\end{gather*}
where $T_2^* = n^{-2}\ve^\top (\mX\mX^\top - \mathbb D_X)\ve, \tilde R_1 = n^{-2} \{\tilde\ve^\top(\mX\mX^\top -\mathbb D_X)\tilde\ve - \ve^\top(\mX\mX^\top - \mathbb D_X)\ve\}, \tilde R_2 = n^{-2}\tilde\ve^\top \widehat\mZ(\widehat \mZ^\top\widetilde \mW\widehat \mZ)^{-1} \widehat \mZ^\top\widetilde\mW(\mX \mX^\top-\mathbb D_X)\widetilde\mW\widehat\mZ(\widehat \mZ^\top\widetilde \mW\widehat \mZ)^{-1} \widehat \mZ^\top\tilde\ve$, and $\tilde R_3 = n^{-2}\tilde\ve^\top(\mX \mX^\top-\mathbb D_X)\times\widetilde\mW\widehat\mZ(\widehat \mZ^\top\widetilde \mW\widehat \mZ)^{-1} \widehat \mZ^\top\tilde\ve$. Then the theorem conclusion follows if we can show the following four results
\begin{gather}
	(1) E(T_2^*) = 0,\var(T_2^*) = 2n^{-2}\tr(\Sigma_\gamma^2)\{1+o(1)\}\label{thm4con1}\\
	(2) \tilde R_1 = o_p\{n^{-1}\sqrt p\},\label{thm4con2} \\
	(3) \tilde R_2 = o_p\{n^{-1}\sqrt p\},\label{thm4con3} \\
	(4) \tilde R_3 = o_p\{n^{-1}\sqrt p\} \label{thm4con4}.
\end{gather}
These four conclusions are counterparts for the conclusion \eqref{thm3con1}-\eqref{thm3con5}. We next show the four results sequentially.

\textsc{Step 1.} By direct computation, we have $E(T_2^*) = E(T_1^*) - n^{-2}E(\ve^\top \mathbb D_X\ve) = n^{-1}\tr(\Sigma_\gamma) - n^{-2}E(\sum_{i=1}^{n}\ve_i^2 X_i^\top X_i) = 0$. Then we consider the variance $\var(T_2^*)= n^{-4}E(T_2^{*2}) = E(\ve^\top\mX\mX^\top \ve)^2 + n^{-4}E(\ve^\top \mathbb D_X\ve)^2 - 2n^{-4}E(\ve^\top \mX\mX^\top \mathbb D_X\ve)$. By the proof in Theorem \ref{Theorem 3}, we have $E(\ve^\top\mX\mX^\top \ve)^2 = nE\{\ve^4(X_i^\top X_i)^2\} + n(n-1)\tr^2(\Sigma_\gamma) + 2n(n-1)\tr(\Sigma_\gamma^2)$. Then we have
\begin{gather*}
	E(\ve^\top \mathbb D_X\ve)^2 = E\bigg\{\Big(\sum_{i=1}^n \ve_i^2 X_i^\top X_i\Big)^2 \bigg\}
	=\sum_{i=1}^{n}E\bigg\{\ve_i^4 \Big(X_i^\top X_i\Big)^2 \bigg\}+ \sum_{i\neq j}E\Big(\ve_i^2\ve_j^2 X_i^\top X_i X_j^\top X_j\Big)\\
	= nE\bigg\{\ve_i^4 \Big(X_i^\top X_i\Big)^2 \bigg\} + n(n-1)E^2\Big(\ve_i^2 X_i^\top X_i\Big) = nE\bigg\{\ve_i^4 \Big(X_i^\top X_i\Big)^2 \bigg\} + n(n-1)\tr^2(\Sigma_\gamma).
\end{gather*}
By straightforward computation, we have
\begin{gather*}
	E(\ve^\top \mX\mX^\top \mathbb D_X\ve ) = E\bigg\{\Big(\sum_{j_1,j_2}^{n} \ve_{j_1}\ve_{j_2}X_{j_1}^\top X_{j_2}\Big)\Big(\sum_{i=1}^{n}\ve_i^2 X_i^\top X_i \Big) \bigg\}\\
	= \sum_{i,j_1,j_2}E\Big(\ve_i^2 \ve_{j_1}\ve_{j_2} X_i^\top X_i X_{j_1}^\top X_{j_2} \Big) = nE\bigg\{\ve_i^4 \Big(X_i^\top X_i\Big)^2 \bigg\} + \sum_{i\neq j}E\Big(\ve_i^2\ve_j^2 X_i^\top X_iX_j^\top X_j\Big)\\
	 = nE\bigg\{\ve_i^4 \Big(X_i^\top X_i\Big)^2 \bigg\} +n(n-1)\tr^2(\Sigma_\gamma).
\end{gather*}
As a consequence, we have $\var(T_2^*) = 2n^{-2}(1-n^{-1})\tr(\Sigma_\gamma^2)$.

\textsc{Step 2.} By a similar procedure of \textsc{Step 3} in Theorem \ref{Theorem 3}, we can show that $\tilde{R}_1 = O_p(p/N^{1/2}) = o_p(n^{-1}p^{1/2})$.

\textsc{Step 3.} By a similar procedure of \textsc{Step 4} in Theorem \ref{Theorem 3}, we can show that $\tilde{R}_2 = o_p(n^{-1}p^{1/2})$.

\textsc{Step 4.} By Cauchy-Schwarz inequality, we have $\tilde R_3 = o_p(n^{-1}p^{1/2})$. Then we complete the theorem proof.

\csection{Proof of Theorem \ref{Theorem 5}}

By the proof in Theorem \ref{Theorem 4}, we only need to prove the asymptotic normality of $T_2^*$.
We next verify this conclusion by the Martingale Central Limit Theorem \citep{hall1980martingale,lan2014testing}. Let $\Delta_{n,i} = 2n^{-2}\sum_{j=1}^{i-1} \ve_i\ve_j X_i^\top X_j$ and $T_{n,r} = \sum_{i=2}^r \Delta_{n,i}$. Then let $\mF_r = \sigma\{(\ve_1,X_1^\top)^\top,\dots,(\ve_r,X_r^\top)^\top \}$ be the $\sigma$-algebra generated by $(\ve_j,X_j)^\top$ for $1\le j\le r$. Then we have $T_{n,r}\in\mF_r$. Note that for $j \le q<i$, we have $E(\ve_i\ve_jX_i^\top X_j|\mF_q) = E(\ve_i X_i^\top)\ve_jX_j = 0$ and for $q<j \le i-1$, we have $E(\ve_i\ve_j X_i^\top X_j|\mF_q) = E(\ve_i\ve_j X_i^\top X_j) = E(\ve_iX_i^\top)E(\ve_j X_j) = 0$. Then for any $q<i$, we have $E(\Delta_{n,i}|\mF_q)  = 0$. Thus, we have $E(T_{n,r}|\mF_q) = T_{n,q}$ for any $q<r$. As a consequence, the sequence $\{T_{n,r}, 2\le r\le n\}$ forms a martingale with respect to $\mF_{r}$. Let $V_{n,i} = E(\Delta_{n,i}^2|\mF_{i-1})$ and $V_n = \sum_{i=2}^n V_{n,i}$. To apply the Martingale Central Limit Theorem \citep{hall1980martingale,lan2014testing}, we then need to show that: (1) $V_n/\{2n^{-2}\tr(\Sigma_\gamma^2)\} \stackrel{p}{\to}1$ and (2)
\beq
\frac{n^{2}}{2\tr(\Sigma_\gamma^2)}\sum_{i=2}^n E\Big[\Delta_{n,i}^2I\Big\{|\Delta_{n,i}|>\eta \sqrt 2n^{-1}\tr^{1/2}(\Sigma_\gamma^2)\Big\}\Big|\mF_{i-1} \Big]\stackrel{p}{\to}0, \label{martingale clt}
\eeq
for any $\eta>0$. We next verify the above two conclusions separately. For $V_n$ we have
\beqrs
V_n &=& \sum_{i=2}^n E\Big(\Delta_{n,i}^2|\mF_{i-1}\Big)  = 4n^{-4}\sum_{i=2}^nE\Big(\sum_{1\le j_1,j_2\le i-1}\ve_i^2 \ve_{j_1}\ve_{j_2}X_{j_1}^\top X_iX_i^\top X_{j_2} \Big|\mF_{i-1}\Big) \\
& =& 4n^{-4}\sum_{i=2}^n\sum_{1\le j_1,j_2\le i-1} \ve_{j_1}\ve_{j_2}X_{j_1}^\top \Sigma_\gamma X_{j_2}\\
& =& 4n^{-4}\sum_{i=2}^n\sum_{j=1}^{i-1} \ve_{j}^2X_j^\top \Sigma_\gamma X_j
+ 8n^{-4}\sum_{i=2}^n\sum_{1\le j_1<j_2\le i-1} \ve_{j_i}\ve_{j_2}X_{j_1}^\top \Sigma_\gamma X_{j_2} = V_{n1} + V_{n2},
\eeqrs
where $
V_{n1} = 4n^{-4}\sum_{i=1}^{n-1}(n-i) \ve_{i}^2X_i^\top \Sigma_\gamma X_i$ and $V_{n2} = 8n^{-4}\sum_{i_1<i_2}^{n-1}(n-i_2) \ve_{i_1}\ve_{i_2} (X_{i_1}^\top \Sigma_\gamma X_{i_2})$.
Then we study $V_{n1}$ and $V_{n2}$ separately. Specifically, we have $E(V_{n1}) = 4n^{-4} \sum_{i=1}^{n-1}(n-i) \tr(\Sigma_\gamma^2) = 2n^{-2}\tr(\Sigma_\gamma^2)\{1+o(1)\}$. By a similar step in Theorem 1, we can show that $\|X_i\|/\sqrt p$ is sub-Gaussian. As a result, we have $E(X_i^\top X_i)^2 = O(p^2)$. Next, by Cauchy-Schwarz inequality and Condition (C3), we have
\begin{gather*}
	\var(V_{n1}) = 16n^{-8} \sum_{i=1}^{n-1} (n-i)^2 \var\Big(\ve_i^2 X_i^\top \Sigma_\gamma X_i \Big) \\
	\le 16n^{-8} \sum_{i=1}^{n-1} (n-i)^2 C_{\max}^2E\Big(X_i^\top X_i \Big)^2 = O(n^{-5} p^2) = o(n^{-4}p^2) .
\end{gather*}
As a result, $\var(V_{n1}) = o\{ \var^2(T_2^*) \}$. This implies that $V_{n1}/\var(T_2^*) \stackrel{p}{\to}1$. Then we consider $V_{n2}$. It can be verify that $E(V_{n2})  = 0$. We then have
\beqrs
\var(V_{n2}) &=& 64n^{-8}\sum_{1\le i_1<i_2\le n-1}(n-i_2)^2 E\Big\{\ve_{i_1}^2\ve_{i_2}^2\Big(X_{i_1}^\top\Sigma_\gamma X_{i_2}\Big)^2 \Big\}\\
&=&64n^{-8}\sum_{1\le i_1<i_2\le n-1}(n-i_2)^2\tr\Big\{E\Big(\ve_{i_1}^2X_{i_1}X_{i_1}^\top \Sigma_\gamma\Big)E\Big(\ve_{i_2}^2X_{i_2}X_{i_2}^\top \Sigma_\gamma\Big) \Big\}\\
&=& 64n^{-8}\sum_{1\le i_1<i_2\le n-1}(n-i_2)^2\tr(\Sigma_\gamma^4)\le 64n^{-4}\tr(\Sigma_\gamma^4).
\eeqrs
By Condition (C3), we know that $\tr(\Sigma_\gamma^4) \le C_{\max}^4 p$. Thus, we have $\var(V_{n2}) = O(n^{-4}p) = o\{\var^2(T_2^*)\}$. We then prove $V_{n2}/\var(T_2^*) \stackrel{p}{\to}0$.

Next we consider the second conclusion of \eqref{martingale clt}. For any $\eta > 0$, by the Chebychev's inequality, we have
\begin{gather*}
	\frac{n^{2}}{2\tr(\Sigma_\gamma^2)}\sum_{i=2}^n E\Big[\Delta_{n,i}^2I\Big\{|\Delta_{n,i}|>\eta \sqrt 2n^{-1}\tr^{1/2}(\Sigma_\gamma^2)\Big\}\Big|\mF_{i-1} \Big]\le\frac{n^4}{4\eta^2\tr^2(\Sigma_\gamma^2)}\sum_{i=2}^n E(\Delta_{n,i}^4).
\end{gather*}
Thus, the conclusion holds if we can show $\sum_{i=2}^nE(\Delta_{n,i}^4) = o(n^{-4}p^2)$. It follows that
\beqrs
&&\sum_{i=2}^nE(\Delta_{n,i}^4) = 16n^{-8}\sum_{i=2}^n \sum_{1\le j_1,j_2,j_3,j_4\le i-1} E\Big(\ve_i^4\ve_{j_1}\ve_{j_2}\ve_{j_3}\ve_{j_4}X_i^\top X_{j_1} X_i^\top X_{j_2}X_i^\top X_{j_3}X_i^\top X_{j_4}\Big)\\	
&=& 16n^{-8} \sum_{i=2}^n \sum_{j=1}^{i-1} E\Big\{\ve_i^4\ve_j^4 \Big(X_i^\top X_j\Big)^4\Big\}
+ 96n^{-8} \sum_{i=2}^n \sum_{1\le j_1\neq j_2\le i-1} E\Big\{\ve_i^4\ve_{j_1}^2\ve_{j_2}^2 \Big(X_i^\top X_{j_1}\Big)^2\Big(X_iX_{j_2}\Big)^2\Big\}.
\eeqrs
By condition (C4), we have $16n^{-8} \sum_{i=2}^n \sum_{j=1}^{i-1} E\big\{\ve_i^4\ve_j^4 (X_i^\top X_j)^4\big\} \le 16n^{-8} \sum_{i=2}^n \sum_{j=1}^{i-1} \\ E\big(X_i^\top X_j\big)^4 = O(n^{-6}p^2)$. Similarly, by Cauchy-Schwarz inequality, we have
\begin{gather*}
	16n^{-8} \sum_{i=2}^n \sum_{1\le j_1\neq j_2\le i-1} E\Big\{\ve_i^4\ve_{j_1}^2\ve_{j_2}^2 \Big(X_i^\top X_{j_1}\Big)^2\Big(X_i^\top X_{j_2}\Big)^2\Big\}\\
	\le  8n^{-8} \sum_{i=2}^n \sum_{1\le j_1\neq j_2\le i-1} \bigg[E\Big\{\ve_i^4 \ve_{j_1}^4 \Big(X_i^\top X_{j_1}\Big)^4\Big\} + E\Big\{\ve_i^4 \ve_{j_2}^4 \Big(X_i^\top X_{j_2}\Big)^4\Big\}\bigg]\\
	\le 16n^{-5} E\Big\{\ve_i^4 \ve_j^4\Big(X_i^\top X_j\Big)^4\Big\} = O\Big(n^{-5}p^2\Big) = o(n^{-4}p^2).
\end{gather*}
Thus, the second conclusion of \eqref{martingale clt} holds. As a consequence, we have verified the asymptotic normality of $T_2^*$. Then by Theorem \ref{Theorem 4} and Slutsky's Theorem, we conclude that $T_3 \stackrel{d}{\to}N(0,1)$.

\csection{Proof of Theorem \ref{Theorem 6}}

By simple computation, we can rewrite $\wh{\tr(\Sigma_\gamma^2)}$ as $\wh{\tr(\Sigma_\gamma^2)} =(1-n^{-1}) (U_{1} + U_{2})$, where $U_1  = \{n(n-1)\}^{-1}\sum_{i\neq j} \ve_i^2\ve_j^2 (X_i^\top X_j)^2$ and $U_2 = \{n(n-1)\}^{-1}\sum_{i\neq j}(\widehat\ve_i^2 \widehat\ve_j^2 - \ve_i^2 \ve_j^2)(X_i^\top X_j)^2$. Note that $\tr(\Sigma_\gamma^2)\le C_{\max}^2 p$ by Condition (C4). Then the conclusion holds if we can show (1) $U_1 / \tr(\Sigma_\gamma^2)\stackrel{p}{\to} 1$ and (2) $U_2 =o_p(p)$.

\textsc{Step 1.} We first evaluate $U_1$. We start with its expectation. By the independence of different samples, we have $E(U_1) = \{n(n-1)\}^{-1}\sum_{i\neq j}E\{\ve_i^2\ve_j^2 (X_i^\top X_j)^2\} = \tr\{E(\ve_i^2\ve_j^2 X_iX_i^\top X_jX_j^\top)\} = \tr\{E^2(\ve_i^2 X_iX_i^\top)\}= \tr(\Sigma_\gamma^2)$. Thus, $E(U_1) = \tr(\Sigma_\gamma^2)$. Then we evaluate the variance of $U_1$. To this end, we apply the Hoeffding decomposition for high-dimensional U-statistics proposed in \cite{zhong2011tests}. Let $V_i = (\ve_i, X_i^\top)^\top$ for $i=1,\dots,n$, and $h(V_i,V_j) = \ve_i^2\ve_j^2 (X_i^\top X_j)^2$ be the kernel function. Then $h_1(V_i) = E\{h(V_i,V_j)|V_i\} = E\{\ve_i^2\ve_j^2 (X_i^\top X_j)^2|V_i\} = \tr\{E(\ve_i^2\ve_j^2X_iX_i^\top X_jX_j^\top |V_i) \} = \tr(\ve_i^2 X_iX_i^\top \Sigma_\gamma ) = \ve_i^2 X_i^\top \Sigma_\gamma X_i$. By Condition (C3) and the sub-Gaussianity of $X_i$, we have $\zeta_1 = \var\{h_1(V_i)\} = \var(\ve_i^2 X_i^\top \Sigma_\gamma X_i) \le E\{\ve_i^4 (X_i^\top \Sigma_\gamma X_i)^2 \} \le C_{\max}^2E(X_i^\top X_i)^2 \le C_{\max}^2p^2$. In the meanwhile by Condition (C4), we know that $\zeta_2 = \var\{h(V_i,V_j)\} = \var(\ve_i^2\ve_j^2 (X_i^\top X_j)^2) \le E(X_i^\top X_j)^4 \le \tau_{\max} p^2$. As a result, we have $\var(U_1) = 2(n-2)\{n(n-1)\}^{-1} \zeta_1 + \{n(n-1)\}^{-1}\zeta_2 = O(n^{-1}p^2)$. By Chebyshev's inequality, for any $\eta>0$, we have $P(|U_1/\tr(\Sigma_\gamma^2) - 1|>\eta) \le \var(U_1) / \{\tr^2(\Sigma_\gamma^2)\eta^2\} = O(n^{-1})$. Thus we have shown the first conclusion.

\textsc{Step 2.} We consider $U_2$. Let $\phi_i = \psi_i\wh Z_i^\top\widehat\gamma + (1-\psi_i)Z_i^\top\gamma$ and $\tilde\phi_i = \tilde\psi_i\wh Z_i^\top\widehat\gamma + (1-\tilde \psi_i)Z_i^\top\gamma$ for some $\psi_i,\tilde \psi_i \in(0,1)$. By the mean value theorem and Cauchy-Schwarz inequality, we have
\beqrs
	|U_2| &=& \Big|\frac{1}{n(n-1)}\sum_{i\neq j} \Big(\widehat\ve_i^2 \widehat\ve_j^2-\ve_i^2 \widehat\ve_j^2+ \ve_i^2 \widehat\ve_j^2- \ve_i^2 \ve_j^2\Big)\Big(X_i^\top X_j\Big)^2\Big|\\
	&\le& \frac{1}{n(n-1)}\sum_{i\neq j}|\widehat\ve_i^2 - \ve_i^2|\widehat\ve_j^2\Big(X_i^\top X_j\Big)^2 + \frac{1}{n(n-1)}\sum_{i\neq j}\ve_i^2 |\widehat\ve_j^2 - \ve_j^2|\Big(X_i^\top X_j\Big)^2\\
	&\le& \frac{2}{n(n-1)}\sum_{i\neq j}\Big|\widehat\ve_i^2 - \ve_i^2\Big|\Big(X_i^\top X_j\Big)^2
	\le \frac{4}{n(n-1)}\sum_{i\neq j}\Big|g\Big(\wh Z_i^\top \widehat\gamma\Big) - g\Big(Z_i^\top\gamma\Big)\Big|\Big(X_i^\top X_j\Big)^2 \\
	&+& \frac{2}{n(n-1)}\sum_{i\neq j}\Big|g^2\Big(\wh Z_i^\top \widehat\gamma\Big) - g^2\Big(Z_i^\top\gamma\Big)\Big|\Big(X_i^\top X_j\Big)^2 \\
	&=&\frac{4}{n(n-1)} \sum_{i\neq j} \Big|g(\phi_i)\Big\{1-g(\phi_i)\Big\}\Big(\wh Z_i^\top\widehat\gamma - Z_i^\top \gamma\Big)\Big|\Big(X_i^\top X_j\Big)^2\\
	&+& \frac{4}{n(n-1)} \sum_{i\neq j} \Big|g^2(\tilde\phi_i)\Big\{1-g(\tilde\phi_i)\Big\}\Big(\wh Z_i^\top\widehat\gamma - Z_i^\top \gamma\Big)\Big|\Big(X_i^\top X_j\Big)^2\\
	&\le& \frac{8}{n(n-1)}\sum_{i\neq j}\Big|\Big(\wh Z_i - Z_i\Big)^\top \widehat\gamma\Big| \Big(X_i^\top X_j\Big)^2  + \frac{8}{n(n-1)}\sum_{i\neq j}\Big|Z_i^\top\Big(\widehat\gamma- \gamma\Big)\Big|\Big(X_i^\top X_j\Big)^2\\
	&\le& 8\bigg\{\|\widehat\gamma\|\Big(n^{-1}\sum_{i=1}^n \|\wh Z_i - Z_i\|^2 \Big)^{1/2} + \|\widehat\gamma - \gamma\| \Big(n^{-1}\sum_{i=1}^n \|Z_i\|^2 \Big)^{1/2}\bigg\} \\
	&\times&\bigg\{\frac{1}{n(n-1)}\sum_{i\neq j} \Big(X_i^\top X_j\Big)^4\bigg\}^{1/2}
\eeqrs

By {\sc Step 1.1} in the proof of Theorem \ref{Theorem 2}, we know that $n^{-1}\sum_{i=1}^n \|\wh Z_i - Z_i\|^2 = O_p(p/N)$. Recall that we have shown $\|\wh\gamma - \gamma\| = O_p(n^{-1/2})$ in Theorem \ref{Theorem 2}. Then we have $\|\wh\gamma\| = O_p(1)$. By Law of Large Numbers, we know that $n^{-1}\sum_{i=1}^n \|Z_i\|^2 \stackrel{p}{\to} E\|Z_i\|^2$. As a result, we have $n^{-1}\sum_{i=1}^n \|Z_i\|^2 = O_p(1)$. By Condition (C4), we have $E[\{n(n-1)\}^{-1} \sum_{i\neq j}(X_i^\top X_j)^4]=E(X_i^\top X_j)^4 \le \tau_{\max}p^2$. As a result, we have $\{n(n-1)\}^{-1} \sum_{i\neq j}(X_i^\top X_j)^4 = O_p(p^2)$. Combining the above results, we have $U_{2} = \{O_p(\sqrt{p/N}) + O_p(n^{-1/2})\}O_p(p) = o_p(p)$. Then we complete the theorem proof.

\end{document}